\documentclass[useAMS,usenatbib,fleqn]{mn2e}
\usepackage{graphicx} 
\usepackage{color}
\usepackage{float}
\usepackage[justification=centering]{caption}
\usepackage{placeins}
\usepackage{multicol}
\usepackage{amsmath}
\usepackage{verbatim}
\usepackage{array}
\usepackage{soul}
\usepackage{booktabs}
\usepackage{threeparttable}
\usepackage{amssymb}
\usepackage{pifont}
\usepackage{array,multirow}
\usepackage{tikz}
\usetikzlibrary{shapes,backgrounds,calc}

\makeatletter
\tikzset{circle split part fill/.style  args={#1,#2}{%
 alias=tmp@name, 
  postaction={%
    insert path={
     \pgfextra{%
     \pgfpointdiff{\pgfpointanchor{\pgf@node@name}{center}}%
                  {\pgfpointanchor{\pgf@node@name}{east}}%
     \pgfmathsetmacro\insiderad{\pgf@x}
      \fill[#1] (\pgf@node@name.base) ([xshift=-\pgflinewidth]\pgf@node@name.east) arc
                          (0:180:\insiderad-\pgflinewidth)--cycle;
      \fill[#2] (\pgf@node@name.base) ([xshift=\pgflinewidth]\pgf@node@name.west)  arc
                           (180:360:\insiderad-\pgflinewidth)--cycle;            
         }}}}}  
 \makeatother  

\def\be{\begin{equation}} 
\def\ee{\end{equation}}

\def\msun{{\Msun}}

\def\gsim{\lower.5ex\hbox{\gtsima}} 
\def\lsim{\lower.5ex\hbox{\ltsima}} \def\gtsima{$\; \buildrel > \over 
\sim \;$} \def\ltsima{$\; \buildrel < \over \sim \;$} \def\prosima{$\; 
\buildrel \propto \over \sim \;$} \def\gsim{\lower.5ex\hbox{\gtsima}} 
\def\lsim{\lower.5ex\hbox{\ltsima}} 
\def\simgt{\lower.5ex\hbox{\gtsima}} 
\def\simlt{\lower.5ex\hbox{\ltsima}} 
\def\simpr{\lower.5ex\hbox{\prosima}} \def\la{\lsim} \def\ga{\gsim} 
  
 \def\gtsima{$\; \buildrel > \over \sim \;$} 
\def\ltsima{$\; \buildrel < \over \sim \;$} 
\def\gsim{\lower.5ex\hbox{\gtsima}} 
\def\lsim{\lower.5ex\hbox{\ltsima}} 
\def\simgt{\lower.5ex\hbox{\gtsima}} 
\def\simlt{\lower.5ex\hbox{\ltsima}} 
\def\simpr{\lower.5ex\hbox{\prosima}} 
\def\la{\lsim} 
\def\ga{\gsim}

\def\msun{\,{\rm \Msun}}

\def\E3{{\cal E}_{\rm g}^{III}}

\def\Msun{\rm M_\odot}

\def\kpc{\rm kpc}
\def\Msun{\rm M_\odot}

\def\M*{M_*}
\def\Z*{Z_*}
\def\L*{L_*}

\tikzstyle{igm} = [rectangle, rounded corners, minimum width=2cm, minimum height=1cm,text centered, draw=black, fill=gray!30]
\tikzstyle{hgas} = [rectangle, rounded corners, minimum width=2cm, minimum height=1cm,text centered, draw=black, fill=red!30]
\tikzstyle{cgas} = [rectangle, rounded corners, minimum width=2cm, minimum height=1cm,text centered, draw=black, fill=blue!30]
\tikzstyle{stars} = [rectangle, rounded corners, minimum width=2cm, minimum height=1cm,text centered, draw=black, fill=yellow!30]
\tikzstyle{bh} = [rectangle, rounded corners, minimum width=2cm, minimum height=1cm,text centered, draw=black, fill=black!80]
\tikzstyle{res} = [rectangle, rounded corners, minimum width=2cm, minimum height=1cm,text centered, draw=black, fill=green!40]
\tikzstyle{box} = [rectangle, rounded corners, minimum width=2.8cm, minimum height=2.8cm,text centered, draw=black]
\tikzstyle{box2} = [rectangle, rounded corners, minimum width=2.8cm, minimum height=1.5cm,text centered, draw=black, fill=gray!30]
\tikzstyle{box1} = [rectangle, rounded corners, minimum width=1.5cm, minimum height=1cm,text centered, draw=none]
\tikzstyle{arrow} = [thick,->,>=stealth]
\tikzstyle{prog1} = [circle, minimum size=1.5cm,text centered, draw=black, fill=black!25]
\tikzstyle{prog2} = [circle, minimum size=2.2cm,text centered, draw=black, fill=black!25]
\tikzstyle{prog3} = [circle, minimum size=1.9cm,text centered, draw=black, fill=black!25]
\tikzstyle{prog4} = [circle, minimum size=2.5cm,text centered, draw=black, fill=black!25]
\tikzstyle{split1} = [circle split, minimum size=1.3cm, line width=0pt, circle split part fill={blue!30,red!30}]
\tikzstyle{split2} = [circle split, minimum size=2cm, circle split part fill={blue!30,red!30}]
\tikzstyle{split3} = [circle split, minimum size=1.7cm, circle split part fill={blue!30,red!30}]
\tikzstyle{split4} = [circle split, minimum size=2.3cm, circle split part fill={blue!30,red!30}]
\tikzstyle{star1} = [star, star points=8, minimum size=1cm, star point height=2mm, fill=yellow]
\tikzstyle{star2} = [star, star points=8, minimum size=1.5cm, star point height=2mm, fill=yellow]
\tikzstyle{star3} = [star, star points=8, minimum size=1.2cm, star point height=2mm, fill=yellow]
\tikzstyle{star4} = [star, star points=8, minimum size=1.8cm, star point height=2mm, fill=yellow]
\tikzstyle{bh1} = [circle, minimum size=0.3cm,text centered, draw=black, fill=black]
\tikzstyle{bh2} = [circle, minimum size=0.6cm,text centered, draw=black, fill=black]
\tikzstyle{bh3} = [circle, minimum size=0.4cm,text centered, draw=black, fill=black]
\tikzstyle{bh4} = [circle, minimum size=0.8cm,text centered, draw=black, fill=black]

\title[High-z BH growth and jets]{Super-Eddington accretion in high-redshift black holes and the emergence of jetted AGN}
\author[Piana et al.]{Olmo Piana$^{1,2}$\thanks{piana@ntnu.edu.tw}, Hung-Yi Pu$^{1,2}$, Kinwah Wu$^{3}$\\ 
\\
$^{1}$ Department of Physics, National Taiwan Normal University, No. 88,  Section 4, Tingzhou Road, Taipei 116, Taiwan, R.O.C.\\
$^{2}$ Centre of Astronomy and Gravitation, National Taiwan Normal University, No. 88,  Section 4, Tingzhou Road, Taipei 116, Taiwan, R.O.C.\\
$^{3}$ Mullard Space Science Laboratory, University College London; Holmbury St. Mary, Dorking, Surrey RH5 6NT, UK.\\
}

\begin{document} 
 
\date{} 

\maketitle

\begin{abstract}
In this work we study the co-evolution of central black holes (BHs) and host galaxies by utilizing an advanced iteration of the DELPHI semi-analytical model of galaxy formation and evolution. Based on dark matter halo merger trees spanning the redshift range from $z=20$ to $z=4$, it now incorporates essential components such as gas heating and cooling, cold and hot BH accretion, jet and radiative AGN feedback. We show how different BH growth models impact quasar and galaxy observables at $z \geq 5$, providing predictions that will help discriminate between super-Eddington and Eddington-limited accretion models: despite being both consistent with observed properties of SMBHs and their host galaxies at $z \sim 5-7$, they become very clearly distinguishable at higher redshift and in the intermediate mass regime. We find that the super-Eddington model, unlike the Eddington-limited scenario, predicts a gap in the BH mass function corresponding to the intermediate-mass range $10^4\ \mathrm{M_\odot} < M_\mathrm{bh} < 10^6\ \mathrm{M_\odot}$. Additionally, it predicts black holes up to two orders of magnitude more massive for the same stellar mass at $z=9$. The resulting \textit{velocity dispersion -- BH mass} relation at $z\geq 5$ is consistent with local measurements, suggesting that its slope and normalisation are independent of redshift. Depending on the Eddington ratio, we also model the emergence of AGN jets, predicting their duty cycle across as a function of BH mass and their potential impact on the observed number density distribution of high-redshift AGN in the hard X-ray band.
\end{abstract}


\section{Introduction}

The conspicuous presence of supermassive black holes (SMBHs) at $z > 6-7$ \citep[see for instance][]{mortlock2011, wu2015, banados2018, matsuoka2018, wang2021, larson2023} implies that their seeds must have grown in mass by up to $7-8$ orders of magnitudes in just a few hundreds million years. This can be explained thanks to massive BH seeds formation mechanisms or very efficient early growth model 
\citep[see][for a review]{volonteri2021}, and possibly a combination of the two. From this perspective, \cite{bromm2003} first proposed the possibility of forming direct-collapse black hole seeds with masses $M_\mathrm{bh} \sim 10^4-10^5 \mathrm{M_\odot}$, but their specific environmental requirements are supposed to make them rare objects, and so far there is no observational evidence of their existence. For this reasons, models of super-Eddington BH growth via radiatively inefficient slim accretion disks are more and more often being regarded as the solution to this conundrum \citep[e.g.][]{pezzulli2016}. 
Simulations have shown that the feedback from mild super-Eddington accretion rates interferes with the accretion flow itself, making the growth process discontinuous and overall rather inefficient \citep{regan2019}, unless the spin of the black hole remains low, reducing the feedback efficiency \citep{lupi2023}. If on the other hand the black hole can enter a regime of hyper-Eddington accretion with $\dot{M}_\mathrm{bh} \gsim 500 \dot{M}_\mathrm{E}$, then a prolonged isothermal steady accretion flow can form, favouring a much faster growth \citep{inayoshi2016, sugimura2017, takeo2018}. Still, observed SMBHs with $M_\mathrm{bh} \gsim 10^7 \mathrm{M_\odot}$ do not seem to show any indication of undergoing such strong accretion episodes 
\citep[e.g.][]{trakhtenbrot2017}, suggesting that these might be characteristic of earlier phases of BH growth. If so, it is reasonable to expect the huge amount of feedback energy released during a hyper-Eddington accretion episode to leave some imprints in the statistical properties of the young host galaxies. 

Indeed, the observed correlations between the central black hole mass and the mass, luminosity and velocity dispersion of the galactic bulge \citep{ferrarese2000, gebhardt2000, marconi2003, gultekin2009, graham2014} hint to a co-evolution between the BH and its host galaxy \citep[see][for a review]{shankar2009, kormendy2013}, and the widespread detection of gas outflows in galaxy with active BHs \citep[e.g.][]{king2015} allow us to think that the mechanism responsible for such co-evolution is the coupling between supernovae (SN) and BH feedback energies and the interstellar medium (ISM). In particular, SN feedback is effective in removing gas from the central regions of the galaxy, especially during the initial phases of galaxy evolution, when the mass is low and the potential well shallow, hence hampering black hole growth \citep{bower2017, lupi2019}. BH feedback, on the other hand, becomes important at higher masses, and can usually take two different forms \citep[see][for comprehensive reviews]{morganti2017, cielo2018}: the radiative (\textit{quasar}) mode is usually associated to high-luminosity AGN with high accretion rates, and powered by radiation emitted from the accretion disk \citep{shakura1973}, with photons that are able to transfer their momentum to the IGM (intergalactic medium) particles. These AGN are usually associated to fast outflows with high velocities ($> 500\ \mathrm{km/s}$), and their bolometric luminosity correlates with the outflow mass and size \citep{fiore2017}, suggesting that the AGN radiative feedback might indeed be the main driver of outflows. The jet (sometimes called \textit{radio}) mode, instead, is supposed to extract energy from the magnetic field of rotating black holes \citep{blandford1977}, and is considered to be dominant in low-power AGN, though it has also been occasionally observed in luminous quasars. To be more precise, it is thought that for Eddington ratios $\lambda_\mathrm{E} = \dot{M}_\mathrm{bh}/\dot{M}_\mathrm{E} \lsim 0.01$ (where $\dot{M}_\mathrm{bh}$ is the black hole accretion rate and $\dot{M}_\mathrm{E}$ is the Eddington accretion rate) the black hole accretion disk is geometrically thick and optically thin, allowing the formation of a weak but steady jet. At higher accretion rates, for $0.01 \lsim \lambda_\mathrm{E} \lsim 1$, the disk enters an unstable phase and the gas inflow towards the black hole becomes time-dependent. If $\lambda_\mathrm{E} \gsim 1$ then we have a strong optically-thick radiation-pressure-driven wind which again thickens the disk into a torus, producing jets which will be more collimated for wider tori \citep[for a review about relativistic jet formation in AGN see][]{blandford2019}. The effect of jets, which are observed in $\approx 10\%$ of all AGN, goes well beyond the boundaries of the galaxies: their radio cavities filled with plasma outgrow the host halo, preventing the intergalactic medium (IGM) in the cluster from cooling down and pushing it outwards, hence disrupting gas accretion onto the galaxy. The importance of this effect on the IGM is still unclear though, as it can be degenerate with that of stellar feedback. In addition, the amount of energy deposited into the ISM by the jets can vary a lot, depending on their collimation and on the clumpiness of the medium \citep{wylezalek2018, wagner2012, mukherjee2016}.

In this work, we use a refined version of the DELPHI cosmological semi-analytic model \citep{dayal2019, piana2021, piana2022} - benchmarked against galaxy and AGN UV luminosity functions, stellar mass density and UV luminosity density at $z>4$ - to probe the observable properties of the high-redshift AGN population in different scenarios \citep[see][for an alternative version of DELPHI]{trebitsch2023}. In particular, in addition to the previous version of DELPHI, we model the mass budgets of the cold and hot gas phases, which in turn fuel cold and hot black hole accretion \citep[e.g.][]{raimundo2017, storchibergmann2019}, AGN radiative and jet-mode feedback, and the galactic gas cycle of outflows and re-accretion onto the galaxy. The goal is to show how we can observationally distinguish between Eddington-limited and super-Eddington accretion models, helping in defining what is the most typical BH growth path. One of our objectives is to predict the emergence of jets at $z>4$, and to assess their impact on the observational properties of AGN at high-redshift. 
More specifically, we will determine the jet duty cycles as a function of black hole mass, and how the the AGN number densities evolve over the redshift and luminosity space. Hard X-rays from distant AGN are less affected by line-of-sight effects, and also at these energy bands the contribution from the host galaxies is generally insignificant. We will therefore compare our results with reference to those by keV X-ray surveys \citep{ueda2003, barger2005, ueda2014, miyaji2015, aird2015}. 


\section{Model}
\label{model}
Galaxies evolve as their host dark matter halos grow. Our semi-analytic model is built on the merger trees of 550 halos with final masses $M_\mathrm{h} = 10^8-10^{13.5} \mathrm{M_\odot}$, whose merger history and mass evolution are followed from $z=20$ to $z=4$ in time steps of 20 Myr (though we will test some of our results also on a merger tree with a time step of 10 Myr; see also \S \ref{sec:model_setup}). The merger tree algorithm follows the one described in \cite{parkinson2008}. Each of the 550 final halos is assigned a number density consistent with the Sheth-Tormen halo mass function \citep[HMF,][]{sheth-tormen1999} at $z=4$, and all of its progenitors along the merger tree are then assigned the same number density as the final halo, so to reproduce the correct HMF at each redshift. The merger tree has a mass resolution of $10^8 \mathrm{M_\odot}$, which constrains the mass with which new halos form along the merger tree. The newly-formed halos are immediately seeded with gas mass, proportionally to the cosmic $\mathrm{\Omega_b/\Omega_m}$ ratio. The starting leaves of the merger trees can also be seeded with black holes: these seeds are assumed to be $10^{3-4}\msun$ direct-collapse black holes (DCBHs) if the virial temperature of the halo is $\gsim 10^4$ K and the Lyman-Werner (LW) background impinging on the halo is $30 J_{21}$ \citep{dayal2017}, where $J_{21}$ is the LW background expressed in units of $10^{-21} {\rm erg\, s^{-1}\, Hz^{-1} \, cm^{-2} \, sr^{-1}}$ \citep{sugimura2014}. Starting leaves at $z>13$ not meeting these criteria are instead seeded with a $150\msun$ stellar black hole, resulting from the collapse of Pop-III stars in primordial mini-halos \citep[SBH; for a review about BH seeds formation channels see for example][]{latif2016, inayoshi2020, volonteri2021}. It is worth to point out that the evolving BH population is fully dominated by descendants of SBHs, since the number density of DCBH seeds is 2-3 orders of magnitudes lower \citep{dayal2019}. In this section we describe our treatment of the evolution of the dark matter and baryonic mass components along the merger trees, from time step to time step.
\vspace{10mm}
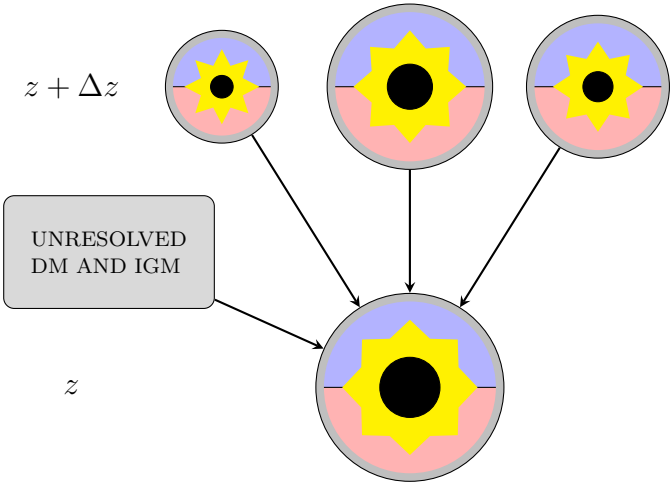
\begin{figure}
\centering
\begin{tikzpicture}[node distance=3cm]

\hspace{-3mm}\node (n0) [box1] {\Large $z+\Delta z$};
\node (n01) [box1, below of=n0, yshift=-10mm] {\Large $z$};
\node (n1) [prog1, right of=n0, xshift=-10mm] {};
\node (n15) [split1, right of=n0, xshift=-10mm] {};
\node (n2) [prog2, right of=n1, xshift=-5mm] {};
\node (n25) [split2, right of=n1, xshift=-5mm] {};
\node (n3) [prog3, right of=n2, xshift=-5mm] {};
\node (n35) [split3, right of=n2, xshift=-5mm] {};
\node (n4) [prog4, below of=n2, yshift=-10mm] {};
\node (n45) [split4, below of=n2, yshift=-10mm] {};
\node (n5) [star1, right of=n0, xshift=-10mm] {};
\node (n6) [star2, right of=n1, xshift=-5mm] {};
\node (n7) [star3, right of=n2, xshift=-5mm] {};
\node (n8) [star4, below of=n2, yshift=-10mm] {};
\node (n9) [bh1, right of=n0, xshift=-10mm] {};
\node (n10) [bh2, right of=n1, xshift=-5mm] {};
\node (n11) [bh3, right of=n2, xshift=-5mm] {};
\node (n12) [bh4, below of=n2, yshift=-10mm] {};
\node (n13) [box2, align=left, below of=n1, xshift=-15mm, yshift=8mm] {UNRESOLVED\\ DM AND IGM};
 
\draw [arrow] (n1) -- (n4);
\draw [arrow] (n2) -- (n4);
\draw [arrow] (n3) -- (n4);
\draw [arrow] (n13) -- (n4);
\end{tikzpicture}
\caption{Schematic view of a dark matter halo merger tree branch: progenitors at $z+\Delta z$ of a halo at $z$ each bring their own contribution of dark matter, (hot and cold) gas, stellar and black hole mass. In addition, smoothly accreted dark matter and gas from the intergalactic space and from the gas reservoir around the halo are accreted according to equations \ref{mdotdm} and \ref{step_ev}.}
\label{merger_tree}
\end{figure}

\subsection{Overview}
\label{model_overview}

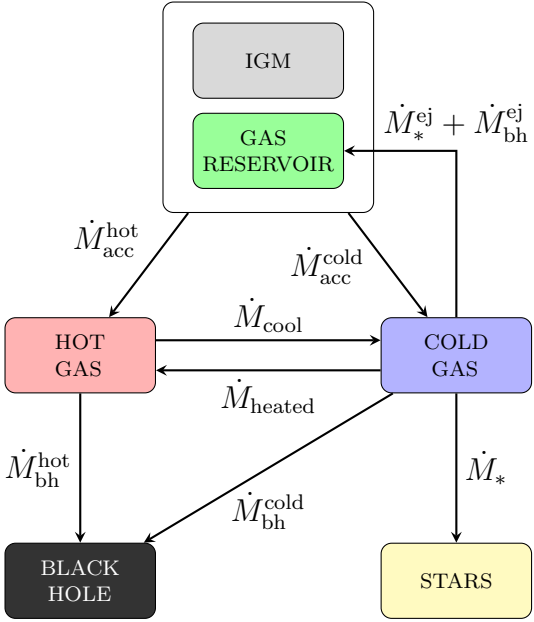
\begin{figure}
\centering
\begin{tikzpicture}[node distance=3cm]
\hspace{0mm} \node (n0) [box] {};
\node (n1) [igm, below of=n0, yshift=36.2mm] {IGM};
\node (n6) [res, align=left, below of=n1, xshift=0mm, yshift=18mm] {\ \ \ \ \ GAS\\ RESERVOIR};
\node (n2) [hgas, align=left, below of=n0, xshift=-25mm, yshift=-3mm] {HOT\\ GAS};
\node (n3) [cgas, align=left, below of=n0, xshift=25mm, yshift=-3mm] {COLD\\ \ GAS};
\node (n4) [bh, align=left, text=white, below of=n2] {BLACK\\ \ HOLE};
\node (n5) [stars, below of=n3] {STARS};
\draw [arrow] (n0) -- (n2) node[midway,above left] {\Large{$\dot{M}_\mathrm{acc}^\mathrm{hot}$}};
\draw [arrow] (n0) -- (n3) node[midway,left=2mm] {\Large{$\dot{M}_\mathrm{acc}^\mathrm{cold}$}};
\draw [arrow] (n2) -- (n4) node[midway,left] {\Large{$\dot{M}_\mathrm{bh}^\mathrm{hot}$}};
\draw [arrow] (n3) -- (n5) node[midway,right] {\Large{$\dot{M}_*$}};
\draw [arrow] (n3) -- (n4) node[midway,below=2mm] {\Large{$\dot{M}_\mathrm{bh}^\mathrm{cold}$}};
\draw [arrow] (n3) |- (n6) node[midway,above] {\Large{$\dot{M}_{*}^\mathrm{ej}+\dot{M}_\mathrm{bh}^\mathrm{ej}$}};
\draw [arrow] ([yshift=2mm] n2.east) -- ([yshift=2mm] n3.west) node[midway,above] {\Large{$\dot{M}_\mathrm{cool}$}};
\draw [arrow] ([yshift=-2mm] n3.west) -- ([yshift=-2mm] n2.east) node[midway,below] {\Large{$\dot{M}_\mathrm{heated}$}};  
\end{tikzpicture}
\caption{Schematic plot of the gas cycle, star formation, and black hole accretion in each dark matter halo at each redshift (the reader can see this figure as the description of what happens in each of the halos in Figure \ref{merger_tree}). See text for the definition of all the terms included here (equations \ref{eq_cold} and \ref{eq_hot}). The hot gas and gas reservoir components represent additions to the model presented in \protect\cite{dayal2019} and \protect\cite{piana2021}.}
\label{gas_cycle}
\end{figure}

Since gas in galaxies represents the fuel for both star formation activity and black hole growth, it is then essential to follow accurately the evolution of the different gas phases to describe galaxy evolution. In order to do so, at each time step we model the mass changes of the following key components: dark matter halo, hot gas, cold gas, black hole, stellar mass, and the halo gas reservoir, in addition to the gas accreted from the IGM. Schematic plots illustrating the relationships between the components are presented in Figures \ref{merger_tree} and \ref{gas_cycle}.

Beside the mass contributions from all of its progenitor, each halo at a time step $z$ will accrete an unresolved amount of dark matter from the intergalactic space according to 
\begin{equation}
M^\mathrm{acc}_\mathrm{dm}(z) = \left[M_\mathrm{h}(z) - \sum_j M_\mathrm{h}^j(z+\Delta z)\right],
\label{mdotdm}
\end{equation}
where the sum runs over all the $j$ progenitors at $z + \Delta z$. Together with the unresolved dark matter, the galaxy accretes from the IGM a gas mass proportional to the cosmic baryonic fraction. The progenitors of a halo will bring in also their content of gas, stars and black hole mass. We can then write  
\begin{equation}
\begin{split}
&M^\mathrm{cold}(z) = \sum_j M_j^\mathrm{cold}(z+\Delta z) + \dot{M}^\mathrm{cold}(z) \tau_s ,\\
&M^\mathrm{hot}(z) = \sum_j M_j^\mathrm{hot}(z+\Delta z) + \dot{M}^\mathrm{hot}(z) \mathrm{\tau_s}, \\
&M_*(z) = \sum_j M^j_*(z+\Delta z) + \dot{M}_*(z) \tau_s,\\
&M_\mathrm{bh}(z) = \sum_j M^j_\mathrm{bh}(z+\Delta z) + \dot{M}_\mathrm{bh}(z) \mathrm{\tau_s},
\end{split}
\label{step_ev}
\end{equation}
where the index $j$ runs over all the progenitors of the halo at the previous time step, and $\tau_s$ is the time step employed in our model, which in our fiducial case is 20 Myr. In the later subsections we will omit the dependence on $z$ from the equations, for simplicity, and we will assume the different terms all refers to the same time step $z$, unless otherwise specified.

\subsection{The gas phases}
\label{gasp} 

The differential change of the cold gas mass within a time step is modelled by: 
\begin{equation}
\begin{split}
\dot{M}^\mathrm{cold} = &\dot{M}^\mathrm{cold}_\mathrm{acc} + \dot{M}_\mathrm{cool} - \dot{M}_* - \dot{M}^\mathrm{cold}_\mathrm{bh} - \dot{M}_*^\mathrm{ej} - \dot{M}_\mathrm{bh}^\mathrm{ej} \\ &- \dot{M}_\mathrm{heated}.
\end{split}
\label{eq_cold}
\end{equation}
In this equation, $\dot{M}^\mathrm{cold}_\mathrm{acc}$ \footnote{Here and in the rest of the paper every dotted quantity is defined as the differential change of that same quantity over one time step. In this case $\dot{M}^\mathrm{cold}_\mathrm{acc} = M^\mathrm{cold}_\mathrm{acc}/\tau_\mathrm{s}$.} is the cold gas accretion rate onto the galaxy, $\dot{M}_\mathrm{cool}$ is the gas cooling rate, $\dot{M}_\mathrm{heated}$ is the gas heating rate because of AGN feedback, $\dot{M}_*$ is the star formation rate, $\dot{M}_\mathrm{bh}^\mathrm{cold}$ is the cold gas mass accretion rate of the black hole, $\dot{M}_*^\mathrm{ej}$ and $\dot{M}_\mathrm{bh}^\mathrm{ej}$ represent the gas ejection rates from the galaxy by SN and AGN feedback respectively. 

Conversely, the change in the amount of hot gas mass is given by
\begin{equation}
\dot{M}^\mathrm{hot} = \dot{M}^\mathrm{hot}_\mathrm{acc} - \dot{M}_\mathrm{cool} - \dot{M}^\mathrm{hot}_\mathrm{bh} + \dot{M}_\mathrm{heated}.
\label{eq_hot}
\end{equation}
Similarly to what we did for the cold gas, $\dot{M}^\mathrm{hot}_\mathrm{acc}$ is the hot gas accretion rate onto the galaxy and $\dot{M}_\mathrm{bh}^\mathrm{hot}$ is the hot gas black hole accretion rate.
From equations \eqref{eq_cold} and \eqref{eq_hot} notice that while we are assuming that stars are formed only from cold ISM (interstellar medium), black holes can accrete both from the hot and cold ISM. In addition, we assume that the effect of SN feedback is solely to drive gas outflows from the host galaxy, while black hole feedback has the potential of both driving gas outflows and heating up part of the cold ISM, depending on whether the jet is active or not.

\vspace{10mm}

The terms $M_\mathrm{acc}^\mathrm{cold}$ and hot $M_\mathrm{acc}^\mathrm{hot}$ represent the cold and hot gas masses accreted from the IGM and from the gas reservoir formed around the halo. Simulations have shown that galaxies will accrete preferentially hot or cold gas depending on their halo mass. If the halo mass $M_\mathrm{h}$ is lower than a critical value $M_\mathrm{h}^\mathrm{crit} \sim 10^{12} \mathrm{M_\odot}$, there is no virial shock surrounding the halo and the gas is accreted from the IGM onto the galaxy in cold mode \citep{keres2005, dekel2006, ocvirk2008}. Above the same critical halo mass, on the other hand, pressure builds up and a virial shock develops, leading to quasi-spherical hot gas accretion mode. In this case though, accretion of cold gas filaments travelling inwards from large scales is still possible \citep[for a review see][]{dayal2019}. In our model, for simplicity, we introduce a parameter $f_\mathrm{cold} = 0.6$ that represents the fraction of gas that is accreted cold. We checked that if we implement a linear halo mass dependence of $f_\mathrm{cold}$ that follows the trend recovered by numerical simulations \citep[see Figure 4 of][]{ocvirk2008}, the change in our results is negligible. in our model, the total gas accretion rate onto the galaxy is defined as 
\begin{equation}
\dot{M}_\mathrm{acc} = (\mathrm{\Omega_b}/\mathrm{\Omega_m}) \dot{M}_\mathrm{dm} + \dot{M}_\mathrm{ret},
\end{equation}
where the term $\dot{M}_\mathrm{ret}$ represents the return rate of coronal gas mass $M_\mathrm{res}$ that falls back onto the galaxy from the reservoir, and reads
\begin{equation}
\dot{M}_\mathrm{ret} = \mathrm{\alpha_{ret}} M_\mathrm{res}/\tau_\mathrm{dyn},
\label{mret}
\end{equation}
with $\tau_\mathrm{dyn} = R_\mathrm{vir}/V_\mathrm{vir}$, being the halo dynamical timescale an $\mathrm{\alpha_{ret}}$ a free parameter. The gas reservoir $M_\mathrm{res}$ is formed by the gas ejected by SN and AGN feedback. We then define
\begin{equation}
\dot{M}^\mathrm{cold}_\mathrm{acc} = \begin{cases}
\dot{M}_\mathrm{acc} & \text{if $M_\mathrm{h} < M_\mathrm{h}^\mathrm{crit}$}\\
f_\mathrm{cold} \dot{M}_\mathrm{acc} & \text{if $M_\mathrm{h} > M_\mathrm{h}^\mathrm{crit}$}
\end{cases}
\label{cold_acc}
\end{equation}
and 
\begin{equation}
\dot{M}^\mathrm{hot}_\mathrm{acc} = \begin{cases}
0 & \text{if $M_\mathrm{h} < M_\mathrm{h}^\mathrm{crit}$}\\
(1-f_\mathrm{cold}) \dot{M}_\mathrm{acc} & \text{if $M_\mathrm{h} > M_\mathrm{h}^\mathrm{crit}$}
\end{cases}
\label{hot_acc}
\end{equation}

\begin{table*}\begin{center}
\begin{threeparttable}
\setlength{\extrarowheight}{3pt}
\caption{Model parameters, default values and defining equation.}
\begin{tabular}{cccc}
\hline
\hline
Parameter & Description & Value & Equation\\
\hline
$f_\mathrm{cold}$ & fraction of the gas mass accreted onto the galaxy as cold & 0.6 & \ref{cold_acc}, \ref{hot_acc}\\
$f_\mathrm{*}$ & star formation efficiency cap & 0.02 & \ref{fstareff}\\
$f_*^\mathrm{w}$ & fraction of SN energy that couples to the gas & 0.1 & \ref{fstarej}\\
$M_\mathrm{h}^\mathrm{crit}$ & critical halo mass for BH growth and cold accretion & $10^{11.25} {\Delta_z}^{-3/8} \mathrm{M_\odot}$ & \ref{mhcrit}\\
$\mathrm{M_{mm}}$ & halo mass ratio defining major mergers & 0.1 & \ref{mmm}\\
$f_\mathrm{av}^\mathrm{bh}$ & fraction of cold gas mass that BH can accrete & 0.001 & \ref{favbh}\\
$f_\mathrm{c}$ & limiting cold gas fraction for quasar accretion & 0.6 & \ref{favbh}\\
$f_\mathrm{qso}^\mathrm{w}$ & fraction of BH energy that couples to the gas & 0.003 & \ref{mbhej}\\
$f_\mathrm{jet}^\mathrm{w}$ & fraction of jet energy that drives outflows & 0.003 & \ref{mbhej}\\
$f_\mathrm{jet}^\mathrm{h}$ & fraction of jet energy that heats up the gas & 0.25 & \ref{mgheated}\\
$\mathrm{\alpha_{ret}}$ & fraction of gas in reservoir that falls back onto the galaxy & 0.1 & \ref{mret}\\
\noalign{\smallskip}
\hline
\hline
\end{tabular}
\label{table_params}
\end{threeparttable}
\end{center}\end{table*}

At each time step $z$ a part of the total hot gas mass in the galaxy is then allowed to cool down at a rate $\dot{M}_\mathrm{cool}$. It is generally assumed that the cooled-down mass corresponds to the gas mass enclosed by the cooling radius, within which the cooling timescale is shorter than the free-fall timescale. We need then to impose a density profile for the hot gas, which we assume to follow an isothermal distribution
\begin{equation}
\rho^\mathrm{hot}(r) = \frac{M^\mathrm{hot}}{4 \pi R_\mathrm{vir} r^2},
\end{equation}
where $R_\mathrm{vir}$ is the virial radius of the dark matter halo, defined according to \cite{barkana-loeb2001} as 
\begin{equation}
R_\mathrm{vir} = 0.784 \left(\mathrm{h}
\frac{M_\mathrm{h}}{10^8 \mathrm{M_\odot}} \mathrm{\frac{{\Omega_m}^z}{\Omega_m}} \frac{18 \pi^2}{\Delta_c}\right)^{1/3}  \left(\frac{10}{1+z}\right)\frac{1}{\mathrm{h}}\;\! \kpc. 
\end{equation} 
Here $\Delta_c = 18 \pi^2 + 82(\mathrm{\Omega_m}(z) - 1) - 39(\mathrm{\Omega_m}(z)-1)^2$.
At each time step we compute the cooling timescale at which the hot diffuse gas is able to cool, defined as
\begin{equation}
\tau_\mathrm{cool}(r) = \frac{3 \mu m_\mathrm{p} k_\mathrm{B} T_\mathrm{vir}}{2 \rho^\mathrm{hot}(r) \Lambda(T_\mathrm{vir}, Z)}
\end{equation}
where $\mu = 0.59$ for a fully ionised primordial gas, $m_\mathrm{p}$ is the proton mass, $k_\mathrm{B}$ the Boltzmann constant, $T_\mathrm{vir}$ the virial temperature of the halo and $\Lambda(T,Z)$ is the cooling function as computed by \cite{sutherland-dopita1993}. In our case we assume all of our galaxies have metallicity $Z = 0.05 Z_\odot$. Once we define the virial velocity of the halo as
\begin{equation}
V_\mathrm{vir} = \left(\frac{\mathrm{G} M_\mathrm{h}}{R_\mathrm{vir}}\right)^{1/2},
\end{equation}
we can compute the corresponding virial temperature 
\begin{equation}
T_\mathrm{vir} = \frac{1}{2} \frac{\mu \mathrm{m}_\mathrm{p} {V_\mathrm{vir}}^2}{k_\mathrm{B}}.  
\end{equation}
The portion of gas that can fuel black hole or stellar growth corresponds to the amount of gas that has enough time to cool and to fall to the centre of the potential well, and that should be contained within both the cooling radius and the free-fall radius.
If we assume that the cooling timescale at the cooling radius is similar to the halo dynamical timescale $\tau_\mathrm{dyn} = R_\mathrm{vir}/V_\mathrm{vir}$, we can derive the cooling radius by equating the two quantities, obtaining
\begin{equation}
r_\mathrm{cool} = \left(\frac{M^\mathrm{hot} \Lambda}{6 \mu \pi \mathrm{m}_\mathrm{p} k_\mathrm{B} T_\mathrm{vir} V_\mathrm{vir}}\right)^{1/2}.
\end{equation} 
Similarly, to estimate the free-fall radius, we can compare the dynamical timescale with the free-fall timescale, defined as 
\begin{equation}
t_\mathrm{ff} = \left(\frac{3 \pi}{32 \mathrm{G} \rho}\right)^{1/2} = \left(\frac{3 \pi f^\mathrm{hot}}{32 \mathrm{G} \rho^\mathrm{hot}} \right)^{1/2},
\end{equation}
where $\rho$ is the total density and $f^\mathrm{hot} = \rho^\mathrm{hot}/\rho$ the mass fraction of the hot gas component. Note that here we are assuming that the hot gas density profile tracks that of the total halo mass. By imposing $\tau_\mathrm{ff} = \tau_\mathrm{dyn}$ we define the corresponding free-fall radius
\begin{equation}
r_\mathrm{ff} = \left(\frac{8 M^\mathrm{hot} \mathrm{G} R_\mathrm{vir}}{3 \pi^2 {V_\mathrm{vir}}^2 f^\mathrm{hot}}\right)^{1/2}    
\end{equation}
The accretion radius, within which all the hot gas cools down and is funnelled towards the disk, corresponds then to 
\begin{equation}
r_\mathrm{acc} = \min \left[r_\mathrm{cool}, r_\mathrm{ff}, R_\mathrm{vir} \right]
\end{equation}
and is evaluated at each time step.
We can then write the instantaneous gas cooling rate
\begin{equation}
\dot{M}_\mathrm{cool} = \frac{1}{2} M^\mathrm{hot} \frac{r_\mathrm{acc} V_\mathrm{vir}}{{R_\mathrm{vir}}^2}.
\label{eq_cooling}
\end{equation}

\subsection{Star formation}

Star formation occurs in molecular clouds, so it arises from the cold phase of the gas in the galaxy. At each time step, after implementing the gas cooling mechanism, a fraction $f^\mathrm{eff}_*$ of the cold gas mass forms new stars, and the feedback of the star-forming activity contributes to photo-evaporate part of the remaining cold gas out of the host galaxy. The effective star formation efficiency $f^\mathrm{eff}_*$ is defined as the minimum value between the star formation efficiency whose corresponding SN-II feedback is enough to expel the rest of the gas from the host galaxy and an upper threshold value $f_\mathrm{*} = 0.02$.
We define $E_\mathrm{SN} = f_*^\mathrm{w} \mathrm{E_{51}} \nu \dot{M}_* \mathrm{\tau_s} = \mathrm{f_*^w {v_s}^2}\;\! \dot{M}_* \mathrm{\tau_s}$ as the energy produced by supernovae each time step. Here, $\mathrm{E_{51} = 10^{51} \mathrm{erg}}$ is the energy imparted onto the ISM by each SN-II explosion and $\mathrm{\nu = [134 M_\odot]^{-1}}$ is the number of SNII per stellar mass formed by a Salpeter initial mass function between 0.1 and 100 $\mathrm{M_\odot}$; $f_*^\mathrm{w}$ is the coupling factor between the SN energy and the gas and $\mathrm{v_s}$ is computed to be $611\ \mathrm{km/s}$. We can then write the star formation rate at time step $z$ as
\begin{equation}
\dot{M}_* = [M^\mathrm{cold} + M^\mathrm{cold}_\mathrm{acc} + M_\mathrm{cool}]f^{\mathrm{eff}}_* / \tau_\mathrm{s}.
\end{equation}
The energy required to unbind the cold gas remaining after the star formation burst is 
\begin{equation}
E_\mathrm{ej} = \frac{1}{2} \left[M^\mathrm{cold} + (\dot{M}^\mathrm{cold}_\mathrm{acc} + \dot{M}_\mathrm{cool} - \dot{M}_*)\mathrm{\tau_s}\right] {v_\mathrm{e}}^2,
\end{equation}
where the escape velocity can be written in terms of the halo rotational velocity as $v_\mathrm{e} = \sqrt{2}\;\!V_\mathrm{vir}$. 
Equating $E_\mathrm{SN}$ and $E_\mathrm{ej}$ we obtain the feedback-limited star formation efficiency 
\begin{equation}
f_*^\mathrm{ej} = \frac{{v_\mathrm{c}}^2}{{v_\mathrm{c}}^2
+f_*^\mathrm{w} {v_\mathrm{s}}^2}.
\label{fstarej}
\end{equation} 
The effective star formation efficiency then reads 
\begin{equation}
f_*^\mathrm{eff} = \min\left[f_\mathrm{*}, f_*^\mathrm{ej}\right],
\label{fstareff}
\end{equation}
while the gas mass ejected by SN feedback in one time step is 
\begin{equation}
M_*^\mathrm{ej} = \left[M_\mathrm{cold} + (\dot{M}_\mathrm{cold}^\mathrm{acc} + \dot{M}_\mathrm{cool} - \dot{M}_*)\mathrm{\tau_s}\right]\left(\frac{f_*^\mathrm{eff}}{f_*^\mathrm{ej}}\right).
\end{equation}


\begin{table*}\begin{center}
\begin{threeparttable}
\setlength{\extrarowheight}{3pt}
\caption{List of models.}
\begin{tabular}{cccc}
\hline
\hline
Model & Gas phases & BH accretion & Merger tree time step\\
\hline
\textit{fid} & hot and cold & super-Eddington & 20 Myr\\
\textit{1phase} & only cold & super-Eddington & 20 Myr\\
\textit{EDDlim} & hot and cold & capped at Eddington rate & 20 Myr\\
\textit{10myr} & hot and cold & super-Eddington & 10 Myr\\
\noalign{\smallskip}
\hline
\hline
\end{tabular}
\label{table_models}
\end{threeparttable}
\end{center}\end{table*}

\subsection{Black hole accretion}
\label{bhacc} 

The central black hole grows through mergers and accretion of both hot and cold gas. Because of its limited physical size, its accretion rate strongly depends on the environment and on the gas supply in the centre of the galaxy. In particular, \cite{bower2017} made the point that SN feedback in low-mass halos are very effective in driving cold gas outflows away from the galactic centre, hence starving the central black hole. In such cases, the cold gas bubbles heated by the SN energy are hotter than the gas in the external regions of the galaxy, and, being buoyant, will travel outwards. In high-mass halos, on the other hand, galaxies have a higher virial temperature, and these same bubbles are not buoyant anymore. The gas remains then in the central region of the galaxies, providing the fuel for black hole growth. They estimated the threshold halo mass above which SN feedback is not effective anymore to be 
\begin{equation}
M_\mathrm{h}^\mathrm{th}(z) \sim 10^{12} {\Delta_z}^{-3/8} \mathrm{M_\odot},
\end{equation}
where $\Delta_z = [\mathrm{\Omega_m}(1+z)^3+\mathrm{\Omega_\lambda}]^{1/3}$. This result is consistent with what is shown also in numerical simulations \citep{rosasguevara2016, lupi2019}. Since what marks the end of the stunted black hole accretion regime, physically speaking, is the rise of the halo virial temperature, and given that the two values are found to be similar, we assume $M_\mathrm{h}^\mathrm{th} \equiv M_\mathrm{h}^\mathrm{crit}$. We treat this critical halo mass as a free parameter of the model, and tune it to be 
\begin{equation}
    M_\mathrm{h}^\mathrm{crit}(z) = 10^{11.25} {\Delta_z}^{-3/8} \mathrm{M_\odot},
    \label{mhcrit}
\end{equation}
and we write the mass accreted by the black hole at each time step as 
\begin{equation}
\dot{M}_\mathrm{bh} = \begin{cases}
\dot{M}_\mathrm{bh}^\mathrm{hot} & \text{if $M_\mathrm{h} < M_\mathrm{h}^\mathrm{crit}$}\\
\dot{M}_\mathrm{bh}^\mathrm{hot}+\dot{M}_\mathrm{bh}^\mathrm{cold} & \text{if $M_\mathrm{h} > M_\mathrm{h}^\mathrm{crit}$}
\end{cases}
\end{equation}
where the inflow of sparse hot gas towards the central black hole is allowed at all times. 

In our model, as in many others, hot gas accretion onto the central black hole takes the form of Bondi-Hoyle-Lyttleton accretion mechanism. In this case we have that the hot gas mass accretion rate of the black hole is
\begin{equation}
\dot{M}_\mathrm{bh}^\mathrm{hot} = 4 \pi \mathrm{G}^2 \frac{{M_\mathrm{bh}}^2 \rho_\mathrm{bh}}{{c_\mathrm{s}}^3},
\label{eq_bondi}
\end{equation}
where $\rho_\mathrm{bh}$ is the density of the gas surrounding the black hole and $c_\mathrm{s}$ is the sound speed, here approximated by the halo virial velocity $V_\mathrm{vir}$ \citep{croton2016}. To find $\rho_\mathrm{bh}$ we use the so-called \textit{maximal cooling flow} model \citep{nulsen2000}, and we equate the sound travel time across a shell of diameter twice the Bondi radius to the local cooling time, obtaining
\begin{equation}
\rho_\mathrm{bh} = \frac{3}{8} \frac{\mathrm{m_p} \mathrm{\mu k_B} T_\mathrm{vir} {V_\mathrm{vir}}^3}{\mathrm{G} M_\mathrm{bh} \Lambda(T_\mathrm{vir}, Z)}.
\end{equation}

Episodes of cold gas accretion onto the black hole instead are assumed to be triggered by major mergers, characterised by a halo mass ratio 
\begin{equation}
M_\mathrm{h,1}/M_\mathrm{h,2} \geq \mathrm{M_{mm}} = 0.1,
\label{mmm}
\end{equation}
and to last until the cold gas mass fraction $m_\mathrm{c} = M_\mathrm{cold}/(M_\mathrm{h}+M_*+M_\mathrm{cold}+M_\mathrm{hot}+M_\mathrm{bh})$ of the new host has decreased below a fraction $f_\mathrm{c}$ of its value at the moment of the merger. For the duration of the accretion episode, the black hole is allowed to accrete a fixed fraction of the total cold gas mass present in the galaxy after the star formation burst has taken place
\begin{equation}
\dot{M}_\mathrm{bh}^\mathrm{cold} = f_\mathrm{av}^\mathrm{bh} \tilde{M}^\mathrm{cold}/\mathrm{\tau_s},\ \ \ \ \mathrm{if} \ m_\mathrm{c} > f_\mathrm{c},
\label{favbh}
\end{equation}
where we defined 
\begin{equation}
\tilde{M}^\mathrm{cold} = M^\mathrm{cold} + [\dot{M}^\mathrm{cold}_\mathrm{acc} + \dot{M}_\mathrm{cool} - \dot{M}_* - \dot{M}_*^\mathrm{ej}]\;\! \mathrm{\tau_s}.
\end{equation} 

\subsection{AGN feedback}

In our model we assume that gas outflows are launched by radiative feedback during cold gas accretion while jets contribute to heat up the cold gas in the galaxy. Hence, at each time step, the ejected gas mass can be written as  
\begin{equation}
M_\mathrm{bh}^\mathrm{ej} = \left(\tilde{M}^\mathrm{cold} - \dot{M}_\mathrm{bh}^\mathrm{cold}\mathrm{\tau_s} \right) \frac{f_\mathrm{qso}^\mathrm{w} E_\mathrm{qso} + f_\mathrm{jet}^\mathrm{w} E_\mathrm{jet}}{E_\mathrm{ej}},
\label{mbhej}
\end{equation}
where $f_\mathrm{qso}^\mathrm{w}$ and $f_\mathrm{jet}^\mathrm{w}$ are coupling constants that describe how much of the quasar and jet energy couples to the gas, and for simplicity are tuned to be equal. To compute the total quasar energy $E_\mathrm{qso} = L_\mathrm{qso} \mathrm{\tau_s}$ emitted in a time step, we employ the solution as computed from simulations of relativistic slim accretion disk of \cite{sadowski2009} and fitted by \cite{madau2014}, which takes into account the spin of the black hole and is applicable also in the case of super-Eddington accretion. In this formulation the Eddington rate is defined as $\dot{M}_\mathrm{E} = 16\;\! L_\mathrm{E}/c^2$, with $L_\mathrm{E}$ being the Eddington luminosity. Given our accretion rate $\dot{M}_\mathrm{bh}$, we can then compute the bolometric luminosity $L_\mathrm{qso}$ emitted by the black hole as 
\begin{equation}
\frac{L_\mathrm{qso}}{L_\mathrm{E}} = A(a) \left[\frac{0.985}{\dot{M}_\mathrm{E}/\dot{M}_\mathrm{bh}+B(a)} + \frac{0.015}{\dot{M}_\mathrm{E}/\dot{M}_\mathrm{bh}+C(a)}\right],
\label{lqso}
\end{equation}
where 
\begin{equation}
\begin{split}    
& \mathrm{A(a) = (0.9663 - 0.9292a)^{-0.5639}},\\
& \mathrm{B(a) = (4.627 - 4.445a)^{-0.5524}},\\
& \mathrm{C(a) = (827.3 - 718.1a)^{-0.7060}}.
\end{split}
\end{equation}
Here $\mathrm{a} = 0.5$ is the dimensionless spin parameter, taken to be equal for all black holes. This fit is shown to yield acceptable residuals with respect to the numerical results within the ranges $0 < \mathrm{a} < 0.998$. Physically, this means that the radiative efficiency will be lower for black holes characterised by higher (super-Eddington) accretion rates and higher spins. 

\begin{figure*}
\includegraphics[width=\textwidth]{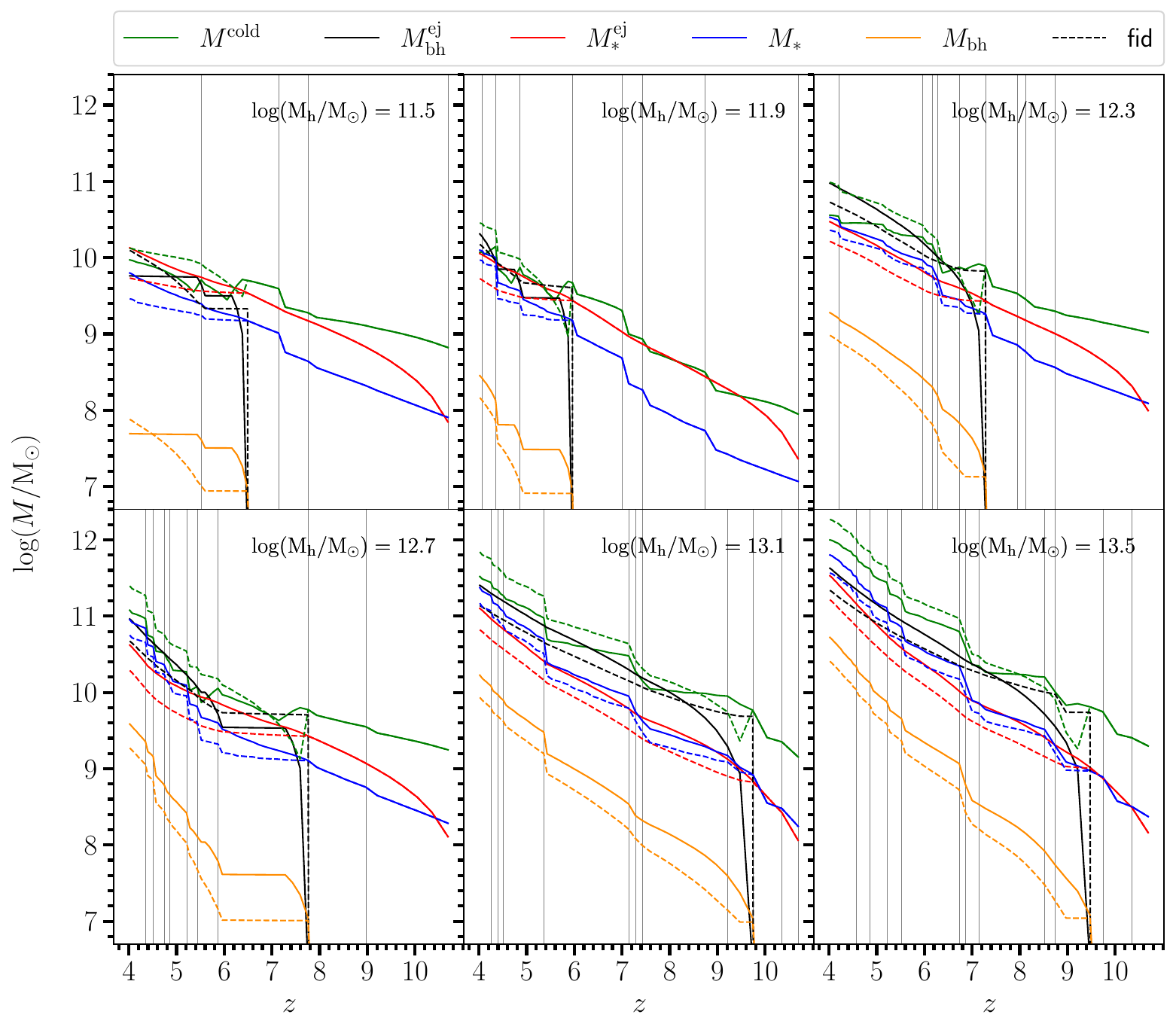}\\
\caption{Mass evolution for different galaxy components for \textit{1phase} model (no gas heating and cooling, solid lines), compared with the \textit{fid} model (dashed lines), here used as a reference. In each panel we follow the growth of the main branch of the merger tree of halos of different mass. We show the total gas mass content ($M^\mathrm{gas}_\mathrm{tot}$) at each time step, the cumulative black hole and stellar mass ($M_\mathrm{bh}$ and $M_*$), the cumulative gas mass ejected by AGN and stellar feedback ($M_\mathrm{bh}^\mathrm{ej}$ and $M_{*}^\mathrm{ej}$). The vertical grey lines indicate all the time steps at which a major merger occurred.}
\label{fig_acch}
\end{figure*}

\begin{figure*}
\includegraphics[width=\textwidth]{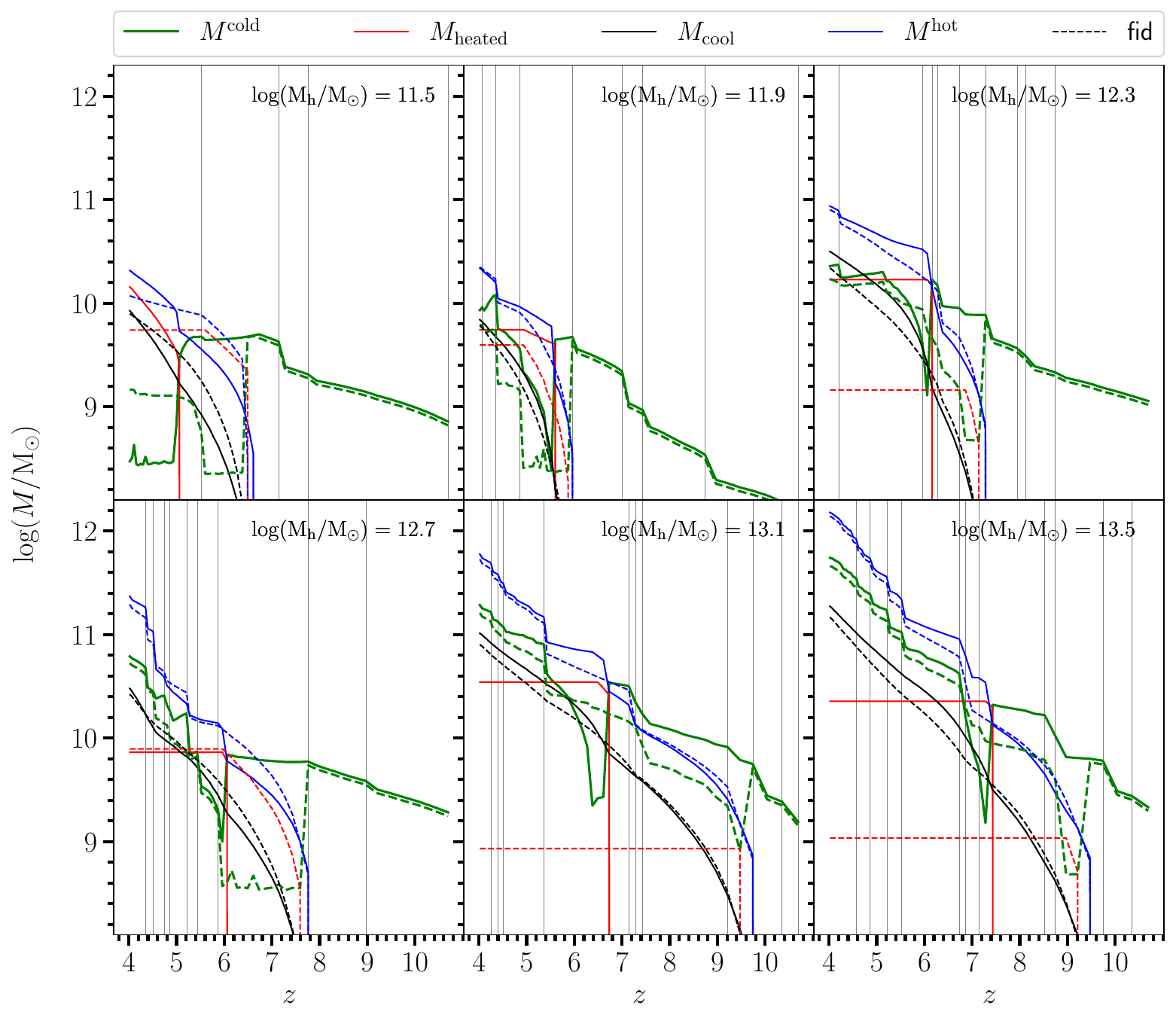}\\
\caption{Mass evolution for different galaxy components for \textit{EDDlim} model (solid lines), compared with the \textit{fid} model (dashed lines), here used as a reference. In each panel we follow the growth of the main branch of the merger tree of the same halos as in Fig.\ \ref{fig_acch}. We show the cold and hot gas mass content ($M^\mathrm{cold}$ and $M^\mathrm{hot}$), and the cumulative heated ($M_\mathrm{heated}$) and cooled ($M_\mathrm{cool}$) gas masses. Notice that the last two quantities and the hot gas mass content are defined only for the \textit{fid} case. The vertical grey lines indicate all the time steps at which a major merger occurred.}
\label{fig_acch_2}
\end{figure*}

Finally, we assume that the jet mode is turned on only when $\lambda_\mathrm{E} \leq 0.01$ or $\lambda_\mathrm{E} \geq 1$, and its power is computed according to the Blandford-Znajek power defined in \cite{tchekhovskoy2011}
\begin{equation}
L_\mathrm{jet} = 2.8\;\! f(a)\left(\frac{\phi}{15}\right)^2 \dot{M}_\mathrm{bh}\;\! c^2,
\end{equation}
where $\phi$ is the dimensionless magnetic flux and $\mathrm{f(a) = a^2(1+\sqrt{1-a^2})^{-2}}$. In this work for each black hole seed we randomly draw $\phi$ from a uniform distribution of values between 1 and 50. Every time there is a black hole merger, the resulting $\phi$ will be that of the black hole with higher mass. Part of the jet energy goes into heating up cold gas present in the galaxy, according to the equation 
\begin{equation}
\dot{M}_\mathrm{heated} = f^\mathrm{h}_\mathrm{jet} \frac{2\;\! L_\mathrm{jet}}{{V_\mathrm{vir}}^2},
\label{mgheated}
\end{equation}
where $f^\mathrm{h}_\mathrm{jet}$ represents the fraction of jet energy that goes into heating up the gas. 

The gas masses ejected in the form of outflows by SN and AGN feedback are accumulated into the gas reservoir ($M_\mathrm{res}$), whose change within a single time step is then computed as  
\begin{equation}
\dot{M}_\mathrm{res} = \dot{M}_*^\mathrm{ej} + \dot{M}_\mathrm{bh}^\mathrm{ej} - \dot{M}_\mathrm{ret}.
\end{equation}


\subsection{Model setup and parameters}
\label{sec:model_setup}
We summarise the free parameters and the adopted values in Table \ref{table_params}. To compare the effects, several different versions of the model are considered, and their differences are outlined in Table \ref{table_models}. In the fiducial ({\it fid}) model, we allow both hot and cold black hole accretion, with no explicit Eddington limit. The dark matter merger trees have a time resolution of 20 Myr. In the single-phase ({\it 1phase}) model, all the gas is assumed to be cold, and no heating mechanisms are considered. Hence, only cold accretion is allowed and all of the feedback energy goes into driving gas outflows. In the Eddington-limited ({\it EDDlim}) model, black hole accretion cannot exceed the Eddington accretion rate. Finally, to explore the time resolution effect of the dark matter merger tree, in the {\it 10myr} model we consider a merger tree with a finer time step, 10 Myr.

\section{Results and Discussions}
\subsection{The importance of including gas heating and cooling}
\label{sec1phase}

\begin{figure*}
\centering
\includegraphics[width=\textwidth]{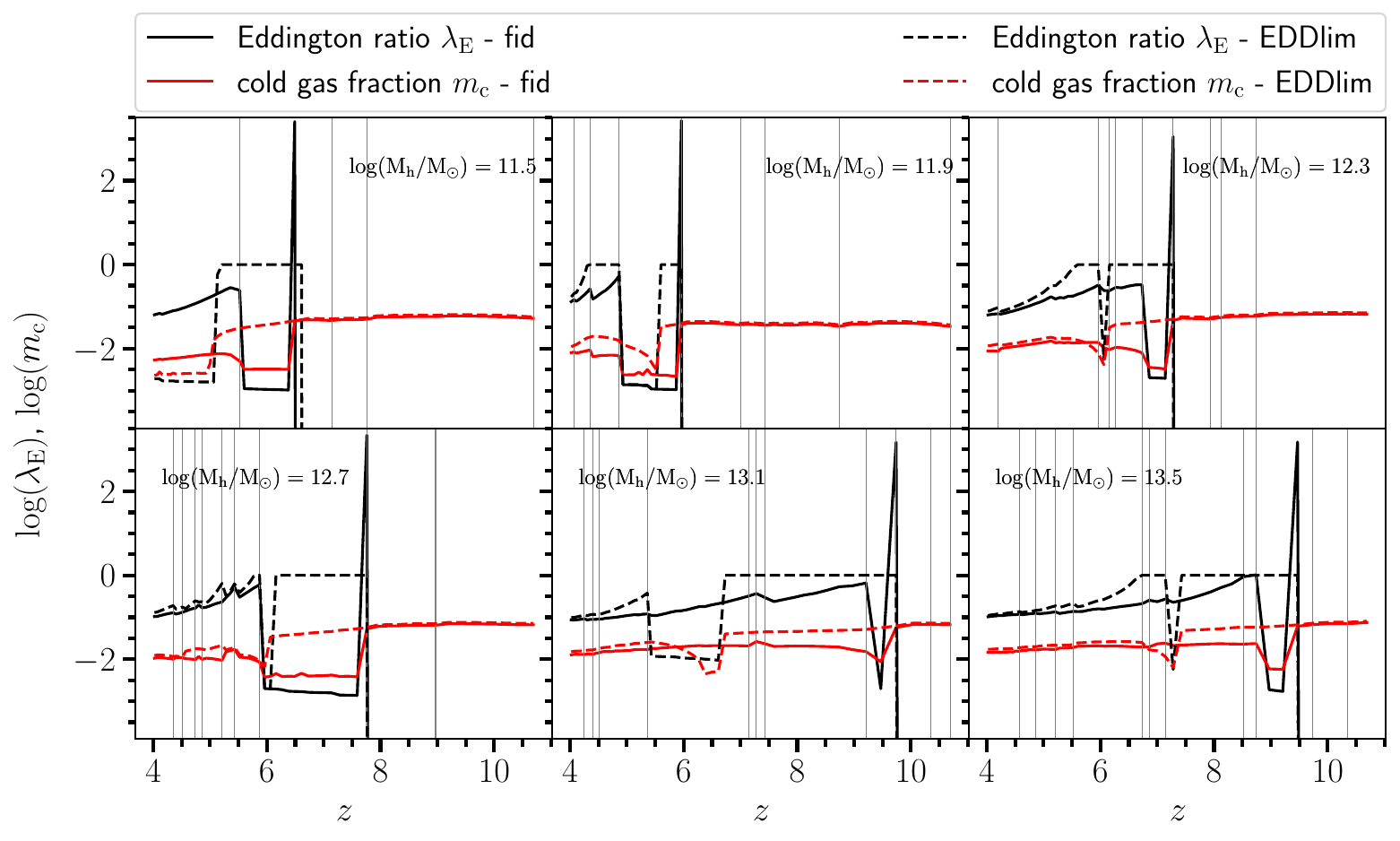}\\
\caption{Evolution of the Eddington ratio $\lambda_\mathrm{E} = \dot{M}_\mathrm{bh}/\dot{M}_\mathrm{E}$ (black lines) and of the cold gas fraction $m_\mathrm{c} = M_\mathrm{cold}/(M_\mathrm{h}+M_*+M_\mathrm{cold}+M_\mathrm{hot}+M_\mathrm{bh})$, (red lines) for the same halos as in Figure \ref{fig_acch}, showing the results for both the \textit{fid} (solid lines) and \textit{EDDlim} (dashed lines) scenarios. The vertical grey lines indicate all time steps at which a major merger occurred.}
\label{fig_eddrat}
\end{figure*}

In order to assess the importance of incorporating into DELPHI cooling and heating processes, we first compare the time evolution of the main branches of selected halos in the \textit{fid} and \textit{1phase} models in Figure \ref{fig_acch}. Starting from the final $z=4$ halo, the main branch is built by selecting at each time step the most massive progenitor. With grey vertical lines, we also show the moments in which the halo undergoes major mergers, which according to our model trigger BH accretion episodes of cold gas (or prolong an ongoing accretion episode) and hence correspond to the start of faster BH growth phases. First of all, we notice that differences between the models arise only when the halo mass crosses the $M_\mathrm{h}^\mathrm{crit}$. In fact in the \textit{fid} model only these higher-mass halos accrete hot gas, below this threshold all of the gas is cold (or assumed to cool down immediately). However, past $M_\mathrm{h}^\mathrm{crit}$ the presence of hot gas in the \textit{fid} model limits the growth of the host galaxy, and as a consequence the \textit{1phase} model can produce more stars and bigger black holes. In turn this corresponds to more gas mass ejected from the galaxy and lower gas mass available for the subsequent time step. Eventually, both black holes and stellar masses turn out to be $\approx 0.2-0.3$ dex higher. Notice that in the \textit{1phase} scenario Bondi accretion starts as soon as the black hole is seeded, while in the \textit{fid} model, where it requires the presence of hot gas, Bondi accretion occurs only for $M_\mathrm{h} > M_\mathrm{h}^\mathrm{crit}$.

\subsection{The impact of super-Eddington accretion}

\begin{figure*}
\includegraphics[width=\textwidth]{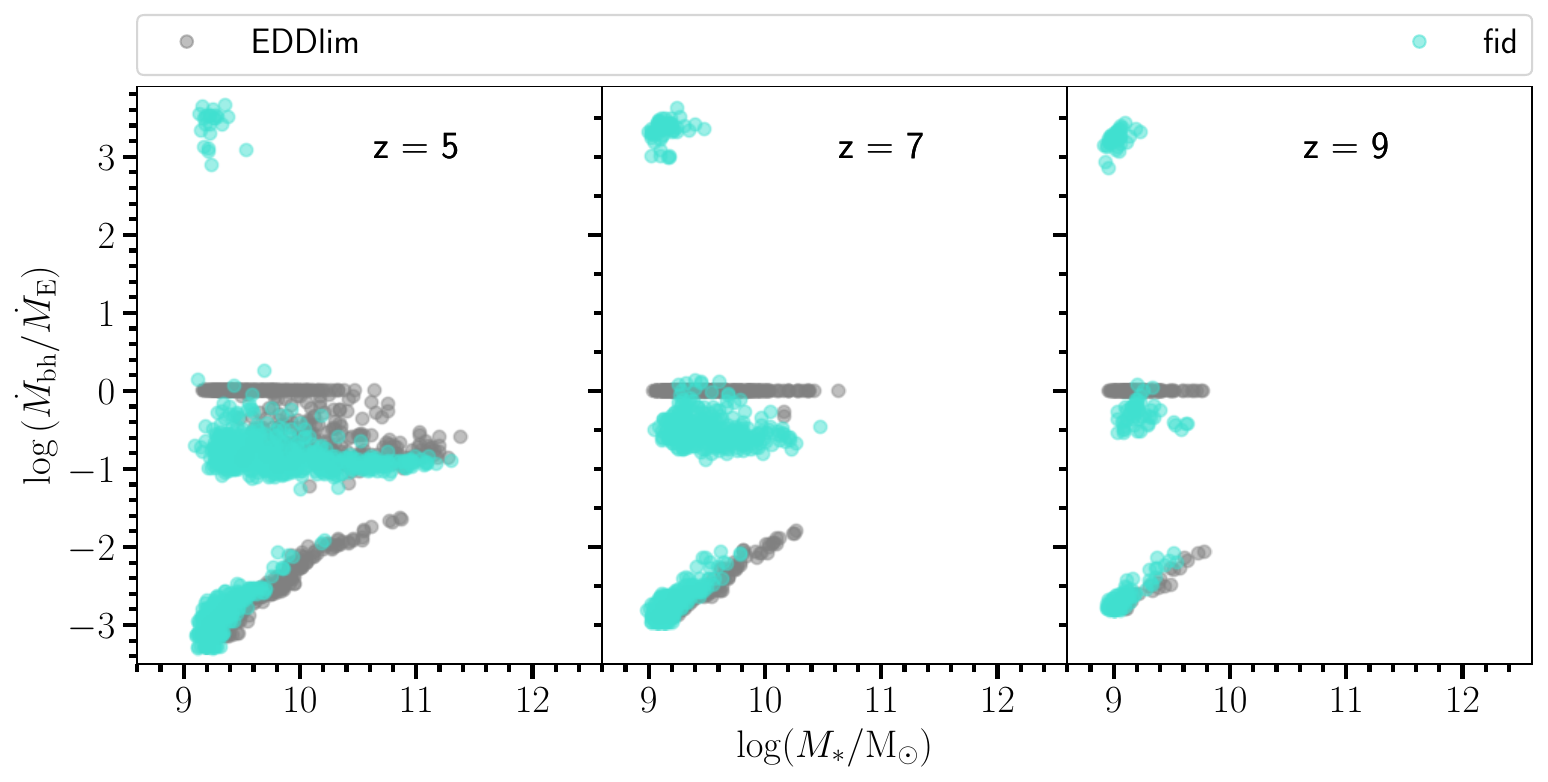}\\
\caption{Galaxy mass-Eddington ratio relation at different redshift. We compare the results from the \textit{fid} (cyan dots) and \textit{EDDlim} (grey dots) models. Notice that even for the highest values of the intrinsic Eddington ratio $\dot{M}_\mathrm{bh}/\dot{M}_\mathrm{E}$, the observed ratio will reach at most values of $L_\mathrm{qso}/L_\mathrm{E} \sim 4$ (eq.\ \ref{lqso}).}
\label{eddr_starm}
\end{figure*}

In Section \ref{sec1phase} we showed that distinguishing between hot and cold gas is important to properly describe the gas cycle of the galaxies. Here, we want to see how the same gas cycle can be affected by different growth models. We do this by comparing the \textit{fid} model, in which BHs are allowed to grow at super-Eddington rate, and the \textit{EDDlim} model, where the BH accretion rate is capped at the Eddington limit. In Figure \ref{fig_acch_2} we plot the evolution of the cold and hot gas masses ($M^\mathrm{cold}$ and $M^\mathrm{hot}$) at  each step, together with the amount of cold gas mass that gets heated up (by AGN feedback, $M_\mathrm{heated}$) and the amount of hot gas that cools down ($M_\mathrm{cool}$). In both models, as soon as the halo outgrows the critical mass ($M_\mathrm{h} > M_\mathrm{h}^\mathrm{crit}$), the hot gas mass dominates the total gas budget, as the galaxy switches from cold to hot-and-cold accretion mode from the IGM. A fraction of the hot gas cools down, and this fraction is lower for higher-mass halos, since from eq.\ \ref{eq_cooling} we can derive that the gas cooling rate follows $\dot{M}_\mathrm{cool} \propto M^\mathrm{hot} r_\mathrm{cool} V_\mathrm{vir} / {R_\mathrm{vir}}^2 \propto {M_\mathrm{h}}^{1/2}$.
It is also interesting to notice that while the amount of gas in outflows increases with halo mass, the total amount of heated gas mass (in \textit{fid}), result of the \textit{jet} phase of the AGN life cycle, does not seem to depend on the mass. This suggests that jet feedback plays a minor role for the host galaxy evolution, and is important only during the first quasar cycle and for halos with $M_\mathrm{h} \sim 10^{11.5}-10^{12.5} \mathrm{M_\odot}$, where it is comparable to the amount of cold gas present in the galaxy. In the \textit{EDDlim} model, the AGN enters its jetted phase only at a later point: this delay allows the galaxy to accumulate up to 50\% more cold gas mass, explaining the relatively higher stellar and black hole masses already found at early times in Fig.\ \ref{fig_acch}. At the same time, once the jetted phase kicks in, given the higher BH mass and higher accretion rate, the AGN is able to heat up up to 10-15 times more gas than in the \textit{fid} model, with an impact on the later evolution of the galaxy. As a consequence, \textit{EDDlim} galaxies will end up containing more hot gas mass than \textit{fid} galaxies, though later growth tends to smooth out all differences.

In Figure \ref{fig_eddrat} we plot the redshift evolution of the Eddington ratio and the cold gas fraction for the same halos as in Figure \ref{fig_acch} for both the \textit{fid} and \textit{EDDlim} models. Again, the vertical lines indicate the occurrence of major mergers. In the first scenario, shortly after crossing the $M_\mathrm{h}^\mathrm{crit}$ threshold, all halos go through a first jetted phase with accretion rates up to $\lambda_\mathrm{E} \sim 10^3$. Subsequent major mergers trigger weaker BH accretion episodes, with $\lambda_\mathrm{E} \sim 0.1-1$, consistently with observation showing that the most massive black holes at $z>4$ mostly grow at sub-Eddington rates \citep{trakhtenbrot2017}. Since the heating feedback occurs only during the jet phase, for $\lambda_\mathrm{E} \leq 0.01$ or $\lambda_\mathrm{E} \geq 1$, the final heated gas mass budget will be dominated by that first hyper-Eddington accretion episode at $M_\mathrm{h} \gsim M_\mathrm{h}^\mathrm{crit}$, which potentially occurs in all halos. This explains why the total heated gas mass in Figure \ref{fig_acch_2} (red lines) appears to be the more or less independent on the final halo mass. In the \textit{EDDlim} scenario, where we do impose an Eddington cap on the accretion rate, the merger-triggered cold accretion episodes alternate with more quiet periods characterised by $\lambda_\mathrm{E} < 0.01$, but by $z=4$ we recover values of Eddington ratio and cold gas fraction very similar to what we find for the \textit{fid} model. This is indicative of the self-regulating action of active black holes: the feedback from bigger black holes impacts the ISM more strongly, slowing down subsequent BH growth. Hence, given different accretion models, if we evolve the galaxy-BH system for long enough we expect differences to be smoothed out due to the action of AGN feedback. 

As we see in Figure \ref{eddr_starm}, in both scenarios our galaxies are divided between those hosting an active black hole accreting both hot and cold gas mass with $\log(\lambda_\mathrm{E}) \gsim -1$, and those hosting quieter black holes accreting only from the hot gas phase with $\log(\lambda_\mathrm{E}) \lsim -1.5$. The separation between these two populations decreases at higher galaxy masses. In fact, the average $\lambda_\mathrm{E}$ of active black holes shows a decreasing trend with mass, as the cold gas supply to the black hole is limited by feedback; at the same time, inactive black holes accrete at Bondi rate $\dot{M}^\mathrm{hot}_\mathrm{bh} \propto M_\mathrm{bh} {M_\mathrm{h}}^{2/3}$ and their average Eddington rate increases with mass. In addition, in the \textit{fid} we see a third smaller population of hyper-Eddington black holes with $\log(\lambda_\mathrm{E}) \gsim 2.5$ at $M_* \sim 10^{9-9.5} \mathrm{M_\odot}$. 

\begin{figure}
\centering
\includegraphics[width=0.5\textwidth]{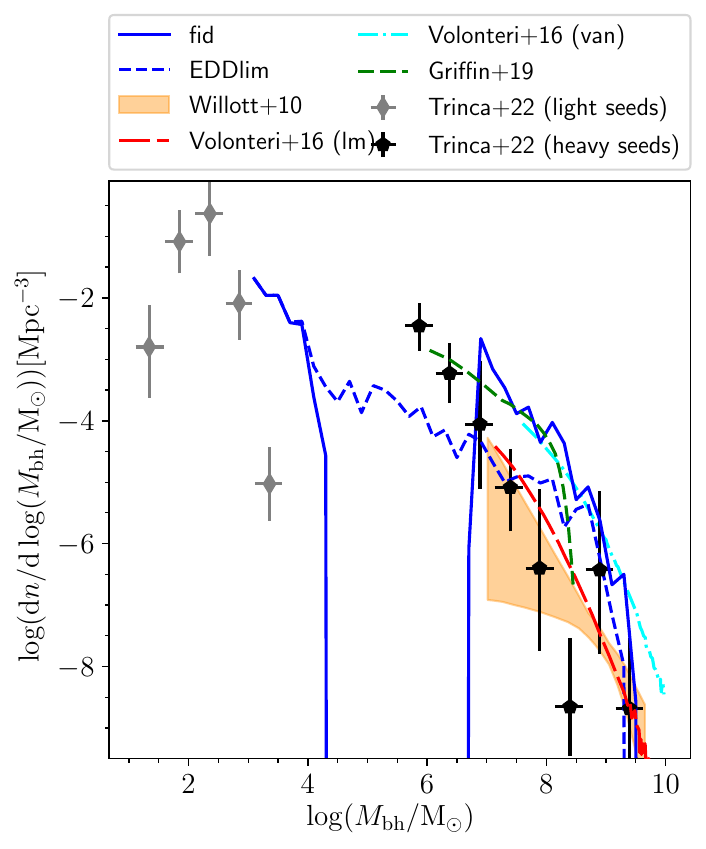}\\
\caption{Total black hole mass function for the \textit{fid} and the \textit{EDDlim} scenarios (solid and dashed blue lines respectively) at $z=6$. We compare our results with those from \protect\citep{volonteri2016a}, who transpose the observed galaxy mass function to the black hole mass function by applying different fits of the $M_\mathrm{bh}-M_*$ relationship, derived from different samples: local galaxies from \protect\cite{marconi2003} and \protect\cite{haring2004} for the vanilla fit (van, dashed-dotted cyan line), low-luminosity AGN in lower-mass galaxies for the low-mass fit (lm, long-dashed red line). We also show the observational results from \protect\cite[][orange shaded area]{willott2010}, as well as results from the \textit{GALFORM} \protect\citep[][dashed dark green line]{griffin2018} and \textit{CAT} \protect\citep[][grey diamonds and black pentagons respectively to indicate descendants of light and heavy BH seeds, born respectively with 
$M_\mathrm{light} \sim 100\;\! \mathrm{M_\odot}$ and $M_\mathrm{heavy} = 10^5 \mathrm{M_\odot}$]{trinca2022} semi-analytic models.}
\label{bhmf}
\end{figure}

One of the consequences of this picture, is that in the \textit{fid} (i.e.\ super-Eddington) model we find the highest values of $\lambda_\mathrm{E}$ for intermediate-mass black holes residing in halos close to $M_\mathrm{h}^\mathrm{crit}$ \citep[see for instance][for a similar conclusion]{ghodla2023}. In this phase, black holes are growing by up to a couple of orders of magnitudes from $M_\mathrm{bh} \sim 10^4 \mathrm{M_\odot}$ to $M_\mathrm{bh} \sim 10^6-10^{6.5} \mathrm{M_\odot}$ within a single time step (i.e.\ 20 Myr), suggesting that the lifetime of intermediate-mass black holes (IMBH) in this scenario is very short, explaining why they still seem to elude detections. This appears clear from Figure \ref{bhmf}, in which we show the evolution of the black hole mass function for both the \textit{fid} and the \textit{EDDlim} case. In the latter, the black hole mass functions is totally depopulated in the mass range $10^4 \mathrm{M_\odot} \lsim M_\mathrm{bh} \lsim 10^6 \mathrm{M_\odot}$. This result should be taken as an indication of the IMBH number density to be very low, rather than actually zero, and exactly how low depends on how smooth the passage from the stunted BH accretion rates in halos with $M_\mathrm{h} < M_\mathrm{h}^\mathrm{crit}$ to the hyper-Eddington regime when $M_\mathrm{h} > M_\mathrm{h}^\mathrm{crit}$. In addition, it is easy to imagine that environmental factors, such as the ionization state of the surrounding IGM, might have an impact on the value of $M_\mathrm{h}^\mathrm{crit}$, which would then not be universal as it is assumed here. It is clear though that in this scenario we still expect the number density of IMBH to be much lower than that of SMBH already at $z=6$. In the \textit{EDDlim} scenario the IMBH number densities at $z=6$ are of the order of $\mathrm{10^{-4}\ Mpc^{-3} dex^{-1}}$, marking a clear difference with respect to \textit{fid}. This difference becomes negligible at the high-mass end of the BHMF though, for $M_\mathrm{bh} \gsim 10^6-10^7 \mathrm{M_\odot}$. Future missions targeting the IMBH mass range at high redshift like \textit{LISA} will be able to put more constraints on the black hole accretion models by observing also the intermediate-mass range of the BHMF. When comparing our SMBH mass function with that from other works, we see that we overpredict the observational constraint set at $z=6$ by \citep{willott2010} by almost one order of magnitude. At the same time we fall in between the two fits proposed by \cite{volonteri2016a}, who derive the BHMF by fitting the observed local $M_\mathrm{bh}-M_*$ for different galaxy samples and applying the fit to the observed galaxy mass function. In addition we reproduce the $z=6$ results from the \textit{GALFORM} semi-analytic model \citep{griffin2018} but with the difference that we can reproduce better the high-mass end of the BHMF for $M_\mathrm{bh}>10^8 \mathrm{M_\odot}$. It is worth to notice that the CAT model \citep[][light green points]{trinca2022} also reproduces a gap in the IMBH mass range, but for a different physical reason: in this case the populations from light and heavy BH seeds occupy two separated regions of the plot.

\begin{figure*}
\includegraphics[width=\textwidth]{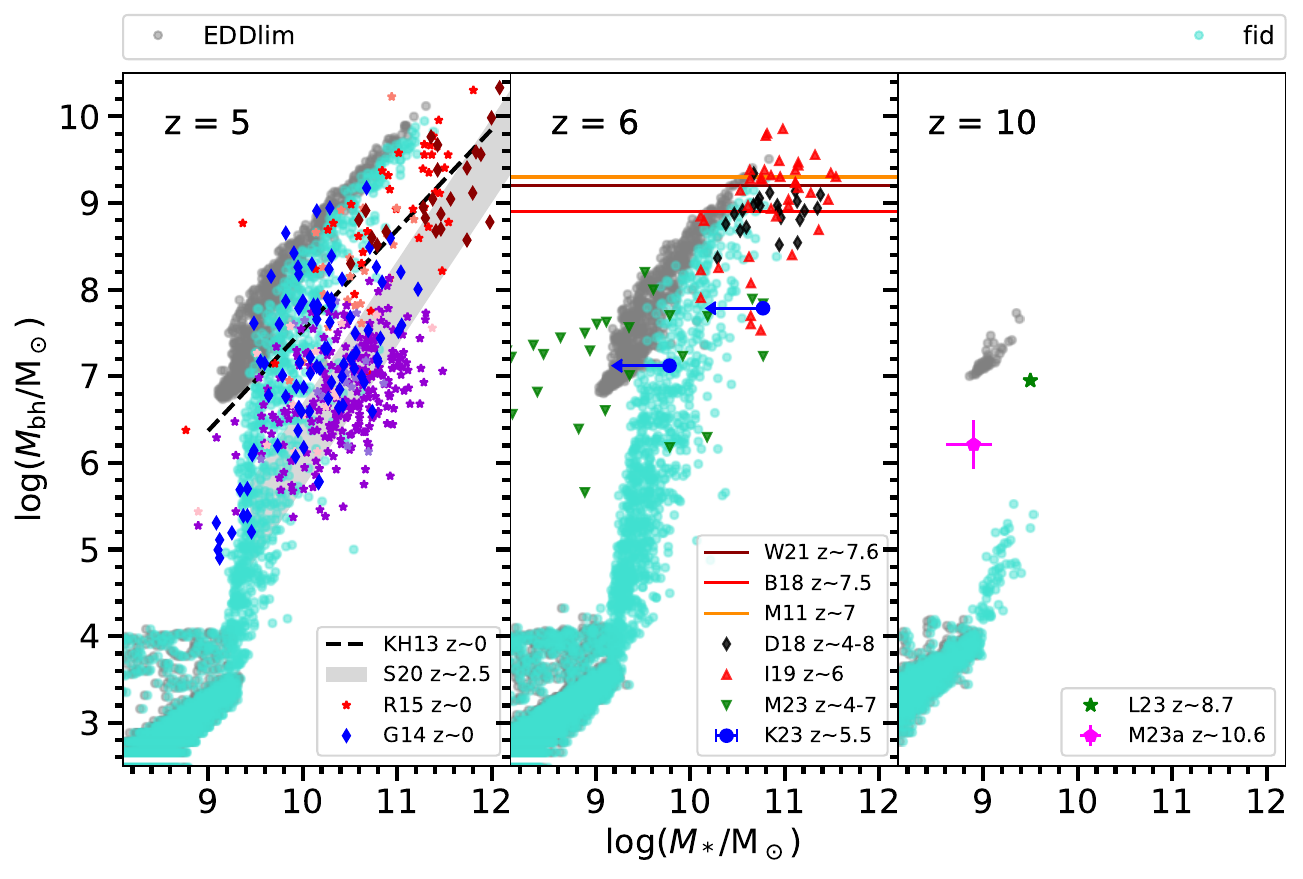}\\
\caption{Galaxy mass-black hole mass relation for all the galaxies in our model at $z = 5, 6$ and 10. We compare the results from the \textit{fid} (cyan dots) and \textit{EDDlim} (grey dots) models. In the z=5 panel we also show $z \sim 0$ observational data from \protect\cite[][G14, blue and red diamonds; direct black hole measurements for blue and red galaxies respectively]{graham2014} and from \protect\cite[][R15, stars]{reines2015} for quiescent elliptical galaxies (red), S/S0 with classical bulges (salmon) and pseudobulges (purple), broad-line low-luminosity AGN (violet), and black holes with reverberation mapping measurements (pink). We also show the $z \lsim 2.5$ correlation derived in \protect\cite[][S20]{suh2020} with its 1-$\sigma$ scatter (grey shaded area) and the one shown in \protect\cite[][KH13, dashed black line]{kormendy2013} for local BH and bulge masses. The horizontal lines in the middle panel correspond to the estimated mass of the highest-$z$ ($z \sim 7$) SMBHs \protect\citep[][respectively M11, W21, B18]{mortlock2011, wang2021,banados2018}. Black diamonds represent ALMA C-II-detected quasars from \protect\cite[][D18]{decarli2018}, red triangles are taken from \protect\cite[][I19]{izumi2019}, green triangles from \protect\cite[][M23]{maiolino2023} and the blue circles from \protect\citep[][K23]{kocevski2023}. In the third panel, together with our $z=10$ results, we show the quasars reported in \protect\cite[][L23, green star]{larson2023} and in \protect\cite[][M23, fuchsia pentagon]{maiolino2023a} at $z \sim 9$ and $z \sim 11$.}
\label{bhm_starm}
\end{figure*}

\subsection{Mass correlation between galaxy and black hole}

From Figures \ref{fig_eddrat} and \ref{bhmf} we can start to see that both individual and statistical observables derived from the \textit{fid} and \textit{EDDlim} models tend to converge towards lower redshift ($z \leq 6$) and higher black hole masses. If we then want to observationally distinguish between the two models we cannot rely on $z<6$ observations of individual sources, but we need to be able to probe a wider population at a higher redshift. With this in mind we plot, in Figure \ref{bhm_starm}, the redshift evolution of the $M_*-M_\mathrm{bh}$ relation for both the \textit{fid} (cyan dots) and \textit{EDDlim} models (grey dots). While at $z=5$ the difference between the two models for $M_\mathrm{bh} \gsim 10^6 \mathrm{M_\odot}$ is negligible, due to the self-regulating BH action we have already seen in Figure \ref{fig_eddrat}, the two high-mass sequences are offset by almost 1 dex at $z=7$, and by almost 2 dex at $z=9$. At these epochs, galaxies with $M_* \sim 10^{9.5} \mathrm{M_\odot}$ in the \textit{fid} and \textit{EDDlim} models respectively host black holes of masses $M_\mathrm{bh} \sim 10^{6.5} \mathrm{M_\odot}$ and $M_\mathrm{bh} \sim 10^8 \mathrm{M_\odot}$. This is indicative of how black hole growth at earlier times proceeds much faster in the \textit{fid} case, but as time goes on the black holes of the \textit{EDDlim} model catch up. A faster early growth corresponds to a greater impact of AGN feedback, resulting in a lower cold gas mass (see Figure \ref{fig_acch}), and hence in a slower subsequent growth. This provides a clear way to distinguish between the two different black hole growth models at $z \gsim 7$. {\it JWST} surveys have already started to populate the $M_*-M_\mathrm{bh}$ \citep[e.g.][]{larson2023, maiolino2023a}, and once the sample of AGN at $z\gsim 7$ reaches a size for sufficient statistical significance in the analysis, we will be able to more firmly establish whether the typical black hole growth model allows for super-Eddington accretion or not.
Looking at the morphology of the relation at $M_\mathrm{h} > M_\mathrm{h}^\mathrm{crit}$, we notice that both of our models show a bend in the slope of the relation at $M_* \sim 10^{9.5} \mathrm{M_\odot}$ and $z \sim 5$, due to the decreasing trend of the Eddington ratio as we move to higher stellar masses (see Figure \ref{eddr_starm}). This is consistent with observational measurements in the local Universe taken from \cite[][star symbols of different shades of red]{reines2015}, who collected data of both quiescent elliptical galaxies and low-luminosity AGN, with the latter falling by more than one order of magnitude below the (tighter) correlations found for massive ellipticals, and characterised by a steeper slope. A mass/morphology-dependent slope is noticeable also if we look at the results from \cite[][red and blue diamonds]{graham2014}, who divided their sample into early-type and late-type galaxies. Despite the big scatter of the inferred relationship, the different slope at high and low masses seems to be a consistent feature, indicating that black holes are not able to accrete as efficiently in higher-mass galaxies. We also show the observational results from \cite[][light grey shaded area]{suh2020} who study the same relationship for a sample of low-luminosity broad-line AGN, finding no significant evolution for $0<z<2.5$. Similar findings have been shown in other works \citep{salviander2015, sun2015}. 
In recent years there have been studies supporting the idea of an evolving normalization of the $M_\mathrm{bh}-M_*$ correlation, with an increasing black hole-to-stellar mass ratio with increasing redshift \citep[see for instance][]{decarli2010, merloni2010, trakhtenbrot2015, caplar2018}. If our $z=5$ results overpredict the normalization of the relationship with respect to local measurements (left panel), they seem to agree better with higher redshift measurements \citep[middle panel][]{decarli2018, izumi2019, maiolino2023, kocevski2023}. However, these comparisons are to be taken carefully, since possible selection bias in the search for black hole hosts at higher redshift might skew the results towards overmassive black holes \citep{lauer2007}, and the completeness level of high-redshift samples is not clear. In our case, the normalization of the relationship evolves (slowly) with $M_\mathrm{h}^\mathrm{crit}(z)$: black holes of the same mass reside in progressively more massive galaxies towards lower redshift.

\begin{figure*}
\includegraphics[width=\textwidth]{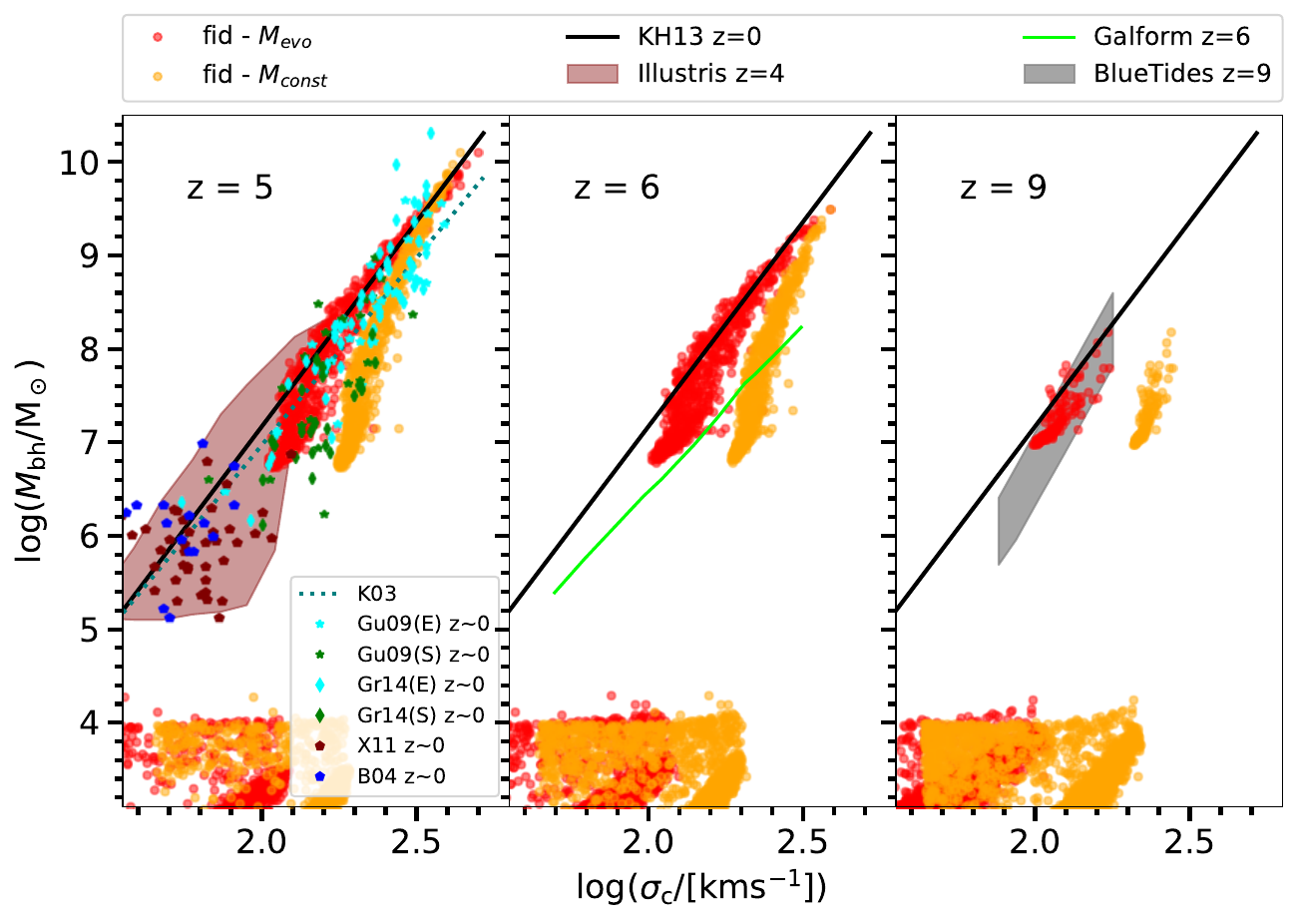}\\
\caption{Central velocity dispersion-back hole mass relation for all the galaxies in our fiducial model at different redshift. $\sigma_\mathrm{c}$ is computed by applying the two fits in \protect\cite{cannarozzo2020}, one assuming an evolving slop ($M_\mathrm{evo}$) and the other one assuming a redshift-independent slope $M_\mathrm{const}$. We also plot the local dynamically-measured black hole catalogues collected in \protect\cite[][Gu09, cyan and green stars respectively for ellipticals and spirals]{gultekin2009} and from \protect\cite[][Gr14, cyan and green diamonds for ellipticals and spirals]{graham2014}, which partially overlap. The (local) low-mass black hole measurements are form \protect\cite[][B09]{barth2004} and \protect\cite[][X11]{xiao2011}. The maroon shaded area represents the $z=4$ black hole population from the Illustris simulation \protect\cite{sijacki2015} while the grey one is the $z=9$ prediction from the BlueTides simulations \protect\cite{huang2019}. The light green solid line is the $z=6$ result from the \textit{Galform} model \protect\citep{malbon2007}, while the dotted line is results from the theoretical argument exposed in \protect\citep[][K03]{king2003}. Finally, as a benchmark, we plot the (local) fit from \protect\citep[][KH13, black line]{kormendy2013}. As we have seen in Figure \ref{bhmf}, our model does not produce intermediate mass black holes at these redshift. At lower redshift, the gap would be populated by black holes accreting at Bondi rate.}
\label{sigma_bhm}
\end{figure*}

Despite the fact that the $M_*-M_\mathrm{bh}$ relationship can generally be used as a SMBH mass predictor in the local Universe, there are instances in which it seems to break down. For example, for red nuggets galaxies, characterised by a relatively small radius with respect to galaxies of the same mass, the $M_*-M_\mathrm{bh}$ relation seems to underestimate the black hole mass. The central velocity dispersion $\sigma_\mathrm{c}$ appears to better predict SMBH masses for these galaxies \citep[e.g.][]{matt2023}. In addition, the $\sigma_\mathrm{c}-M_\mathrm{bh}$ relation exhibits a smaller scatter in the local Universe \citep[for instance][]{kormendy2013}, suggesting that $\sigma_\mathrm{c}$ is a direct tracer of the host dark matter halo mass \cite{ferrarese2002, zahid2018}, which in turn ultimately regulates the amount of gas that fuels both black hole and star formation activities. This motivates a closer look at the evolution of this relationship. Here, we do not directly model the value of the central velocity dispersion $\sigma_\mathrm{c}$ in our galaxies, so in order to estimate it we rely on the $M_*-\sigma_\mathrm{c}$ fits provided by observational studies. Many suggest that the $M_*-\sigma_\mathrm{c}$ relation has a mild but non-negligible redshift evolution, at least for $z>1$. \cite{cannarozzo2020}, based on an extended sample of early-type galaxies (ETGs) at $0<z<2.5$, successfully fit the $M_*-\sigma_\mathrm{c}$ correlation with two different models: the first one, $M_\mathrm{evo}$, with an evolving slope, is parameterised as 
\begin{equation}
\log\left(\frac{\sigma_\mathrm{c}}{\mathrm{km/s}}\right) = \mathrm{\alpha_1}(z)\;\! \log\left(\frac{M_*}{10^{11} \mathrm{M_\odot}}\right) + \mathrm{\beta_1}(z),  
\label{mevo}
\end{equation}
with $\mathrm{\alpha_1}(z) = 0.17+0.18 \log(1+z)$ and $\mathrm{\beta_1}(z) = + 0.51 \log(1+z) + 2.21$.
The other one, $M_\mathrm{const}$, with a constant slope, reads
\label{mnonevo}
\begin{equation}
\log\left(\frac{\sigma_\mathrm{c}}{\mathrm{km/s}}\right) = \alpha_2 \;\! \log\left(\frac{M_*}{10^{11} \mathrm{M_\odot}}\right)+\beta_2(z).      
\end{equation}
with $\mathrm{\alpha_2 = 0.18}$ and $\mathrm{\beta_2}(z) = 0.48 \log(1+z) + 2.21$.
We take our $M_*-M_\mathrm{bh}$ (Figure \ref{bhm_starm}), we compute the velocity dispersion from our stellar masses with both fits (red dots for eq.\ \ref{mevo} and orange for eq.\ \ref{mnonevo}), and we plot then the $\sigma_\mathrm{c}-M_\mathrm{bh}$ relation for our fiducial model in Figure \ref{sigma_bhm}. However, it is important to specify that since the fitted sample is formed by ETGs with $M_* \gsim 3 \times 10^{10} \mathrm{M_\odot}$, our predictions would be meaningful only for the most massive galaxies, assuming these to be the direct progenitors of ETGs at $z \lsim 2.5$. We also show the theoretical predictions from the \textit{Illustris} and BlueTides numerical simulations \citep{sijacki2015, huang2019} and the \textit{GALFORM} semi-analytical model \citep{malbon2007}. While our results match very well with the prediction from BlueTides at $z=9$, we are not able to reproduce the lower-mass BH population of \textit{Illustris}, shown for $z=4$. The same is true for the $z=6$ \textit{GALFORM} results \citep{malbon2007}, that lie slightly below ours, as they predict lower-mass black holes residing in galaxies with the same $\sigma_\mathrm{c}$ at higher redshift. However, our black holes grow bigger, and can reproduce well the high-$z$ quasar population found at $z>5$ (see Figure \ref{bhm_starm}). For comparison we also plot the dynamically-measured black hole catalogues from the local Universe collected in \cite{gultekin2009}, \cite{kormendy2013} and \cite{graham2013}. We find that at $z=5-9$ our results of the $\sigma_\mathrm{c}-M_\mathrm{bh}$ relation from the $M_\mathrm{evo}$ model follow the same trend set by local observations, and they seem to indicate that the $\sigma_\mathrm{c}-M_\mathrm{bh}$ correlation is indeed independent on redshift. Given the different redshift of the sample from which the assumed $M_*-\sigma$ scaling, no strong conclusion can be drawn from a comparison with our results. The theoretical derivation of the $\sigma_\mathrm{c}-M_\mathrm{bh}$ relation from \cite{king2003} lies extremely close to these local observational results, and is also independent on redshift. We also plot the low-mass samples presented in \cite{barth2004} and \cite{xiao2011} down to $M_\mathrm{bh} \sim \mathrm{10^5 M_\odot}$, also observed at $z=0$. Our \textit{fid} model does not produce these intermediate-mass black holes, due to the fact that we consider a universal definition of $M_\mathrm{h}^\mathrm{crit}$. The mass range corresponding to $M_\mathrm{bh} \sim 10^5-10^6 \mathrm{M_\odot}$ could in fact be populated by black holes hosted by halos for which $M_\mathrm{h}^\mathrm{crit}$ is lower than the universal value assumed in this work. In any case, this should not affect the conclusions drawn for SMBH.

\subsection{The jet duty cycle}

\begin{figure}
\includegraphics[width=0.5\textwidth]{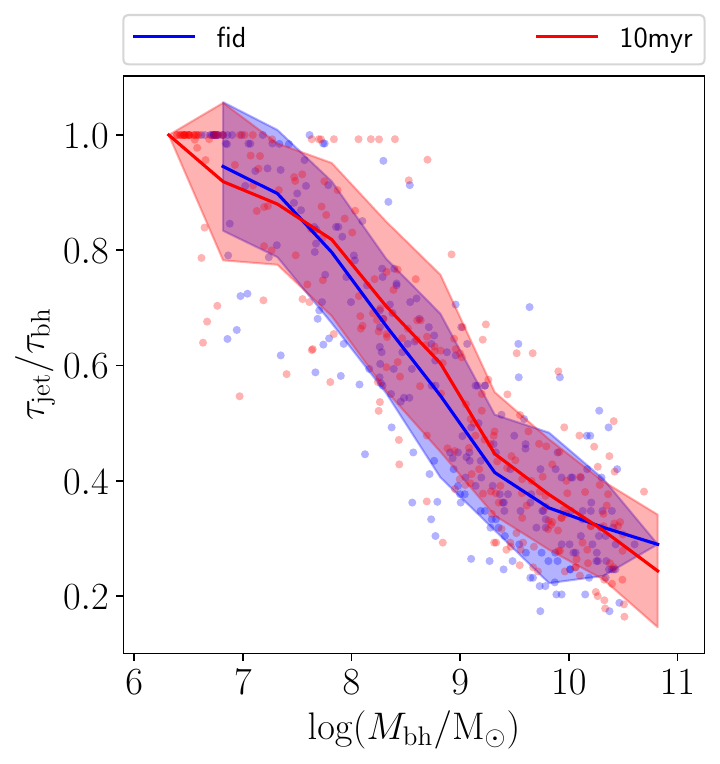}\\
\caption{Fraction of lifetime BHs spend in the jetted phase as a function of black hole mass for $M_\mathrm{bh} > 10^6 \mathrm{M_\odot}$ at $z=4$. We compare the results of our fiducial (blue line) to those of the \textit{10myr} (red line) scenario, in which the time step is $\mathrm{\tau_s} = 10$ Myr, as opposed to $\mathrm{\tau_s} = 20$ Myr for the fiducial case. The line are the averages of the scatter points computed in bins of 0.5 dex in black hole mass, and the shaded areas represent the 1-$\sigma$ deviations from the average.}
\label{sdt}
\end{figure}
Given our jet model, we are able to compute the cumulative jet duty cycle (i.e.\ how much of their total lifetime BHs spend in the jetted phase) as a function of final black hole mass. This is defined here as the intrinsic fraction of lifetime a central black hole spends in the jetted phase since the seed is born. Observationally speaking, the duty cycle is often estimated by computing the ratio of the observed jetted sources over the total number of AGN. In Figure \ref{sdt} we show the average $\mathrm{\tau_{jet}/\tau_{bh}}$ in each BH mass bin of 0.5 dex for all the SMBH in the main branches of our merger trees in the \textit{fid} model, together with its 1-$\sigma$ deviation (blue points, line and shaded area). In order to assess if the outcome depends on the time resolution of the model, we compare it with the results from the \textit{10myr} model (red points, line and shaded area), built exactly like the \textit{fid} model except for the fact that in this case we use a merger tree built with a time step of 10 Myr instead of 20 Myr. Despite the jet fractional lifetimes of each individual black hole can change in the two models, we notice that their statistical distributions show a remarkably precise overlap. In particular, the average jet duty cycle declines from $\mathrm{\tau_{jet}/\tau_{bh} \sim 1}$ from SMBH with final mass $M_\mathrm{bh} \sim 10^6 \mathrm{M_\odot}$ to $\mathrm{\tau_{jet}/\tau_{bh} \sim 0.2}$ for $M_{bh} \sim 10^{11} \mathrm{M_\odot}$. This is consistent with the picture we have drawn in the previous sections, according to which black holes will go through a jetted super-Eddington growth phase early in their lifetime, and then their Eddington ratio will slowly decrease as they grow in mass. 

\subsection{AGN number densities at $z > 5$}
\label{sec_lf}

Observations of absolute AGN number densities, when compared to results from theoretical models, have to rely on obscuration models, which might be different for different observational samples and observed spectral bands. Also for this reason, high-redshift AGN surveys observe in the X-ray band, which is less subject to contamination from the host galaxy and less affected by obscuration effects than other bands. However, it is not clear how much of the total observed X-ray luminosity is coming from the central source and how much of it is contributed by the jets. In fact, if at low redshift relativistic electrons in  jets mostly cool off emitting synchrotron radiation in the radio band, at high redshift the higher CMB energy density $\mathrm{U_{CMB}} \propto (1+z)^4$ means that these same electrons cool off through inverse-Compton scattering against CMB photons. However, we do not expect the particle density and magnetic field to be homogeneous across the jet lobes, and the high-energy regions of the lobes (hot spots) will still be dominated by synchrotron emissions. This might affect the number count of observed high-redshift jetted radio and X-ray sources (in misaligned jets the presence of hot spots might result in additional detection of radio sources) and the total X-ray luminosity \citep{worrall2009, ghisellini2013, ghisellini2014, fabian2014}. Given these uncertainties, we model the hard X-ray luminosity as a combination of the X-ray emission coming from black hole accretion plus a contribution from the jets $L_\mathrm{X,[2-10\ keV]} = L_\mathrm{qso}/\mathrm{K_{X,[2-10keV]}} + \mathrm{\alpha_X} L_\mathrm{jet}$. Here, $\mathrm{K_{X,[2-10keV]}}$ is the hard X-ray bolometric correction derived in \cite{duras2020} for a wide range of luminosities and redshift. At each time step we then select all of the AGN with $L_\mathrm{X} > 10^{41}$ erg/s and we divide them up into 1-dex-wide bins of $L_\mathrm{X}$ and add up their number densities to compute the XLF.

\begin{figure*}
\includegraphics[width=\textwidth]{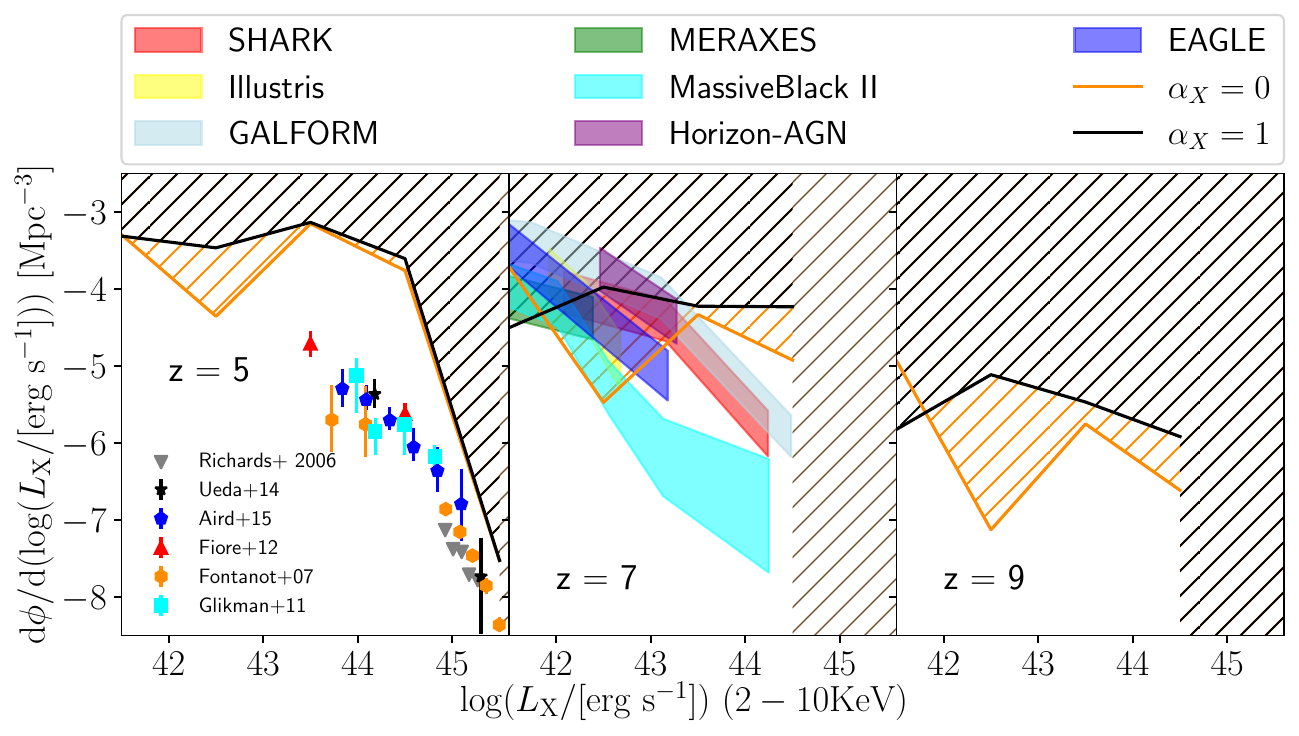}\\
\caption{Intrinsic AGN X-ray luminosity function for the $\mathrm{\alpha_X} = 0$ and $\mathrm{\alpha_X} = 1$ cases. Our results can be considered as upper limits of the observed XLF, as we know that a fraction of AGN are obscured by the central torus, which absorbs part of the light emitted by the black hole \protect\cite{ueda2014, merloni2014}. In the left panel we show our $z=5$ results together with observations of the hard XLF at $z=4-5$ taken from \protect\cite{richards2006}, \protect\cite{fiore2012} and \protect\cite{ueda2014}, at $z=4-5.2$ from \protect\cite{fontanot2007}, at $z = 3.8-5.2$ from \protect\cite{glikman2011}, and at $z=3.5-5$ from \protect\cite{aird2015}. The hatched regions define the region of space forbidden by our model. In the middle panel we compare our $z=7$ results with those from other theoretical models and simulations at $7 < z < 8$, taken from Figure 3 of \protect\cite{amarantidis2019}. In the right panel we show our results at $z=9$. The shaded areas are defined by the corrected and the intrinsic XLF.}
\label{xlf}
\end{figure*}

In Figure \ref{xlf} we show our intrinsic XLF in both cases $\mathrm{\alpha_X} = 0$ (orange line) and $\mathrm{\alpha_X} = 1$ (black line) at $z=5-9$. These are upper limits to the observed XLF, since we are not taking into account torus or dust obscuration models \citep{ueda2014, merloni2014}. Since the fraction of obscured AGN is higher at lower luminosities, our results overpredict the observed XLF especially at lower and intermediate luminosities, with a gap of $\sim 1.5$ dex at $L_\mathrm{X,[2-10\ keV]} \sim 10^{43}-10^{44}$ erg/s, while they are in much better agreement at high luminosities. Keeping into account that obscuration models generally produce an obscured fraction of $\sim 90\%$ at $L_\mathrm{X,[2-10\ keV]} \sim 10^{43}$ erg/s and of $10-15\%$ at $L_\mathrm{X,[2-10\ keV]} \gsim 10^{45}$ erg/s \citep{ueda2014}, our results still lie above the expected XLF by $\sim 0.5$ mag. At $z=7$, we show also the results from several other theoretical models and simulations, re-adapting a Figure taken from \cite{amarantidis2019}, where the shaded areas are defined by the modelled intrinsic and corrected XLF. In this case, our upper limits fall well within the range of predictions from the other models. Nevertheless, we generate a flatter XLF, due to the fact that in our \textit{fid} model SMBH grow very fast early epochs, thanks to an early hyper-Eddington accretion phase, while most of other models implement an Eddington-limited accretion mechanism, making black hole growth more regular throughout time. In addition, in the $\mathrm{\alpha_X} = 0$ model we see a dip in the XLF at $L_\mathrm{X,[2-10\ keV]} \sim 10^{42}-10^{43}$ erg/s, corresponding to a lack of intermediate-mass black holes (see Figure \ref{bhmf}). The dip does not show up if we allow the jet power to contribute to the total X-ray luminosity ($\mathrm{\alpha_X} = 1$), as in this case the black holes accreting at $\lambda_\mathrm{E} < 0.01$ are also forming jets, which give an additional contribution to the X-ray luminosity, pushing the corresponding black holes towards higher luminosity bins. 

\begin{figure}
\includegraphics[width=0.5\textwidth]{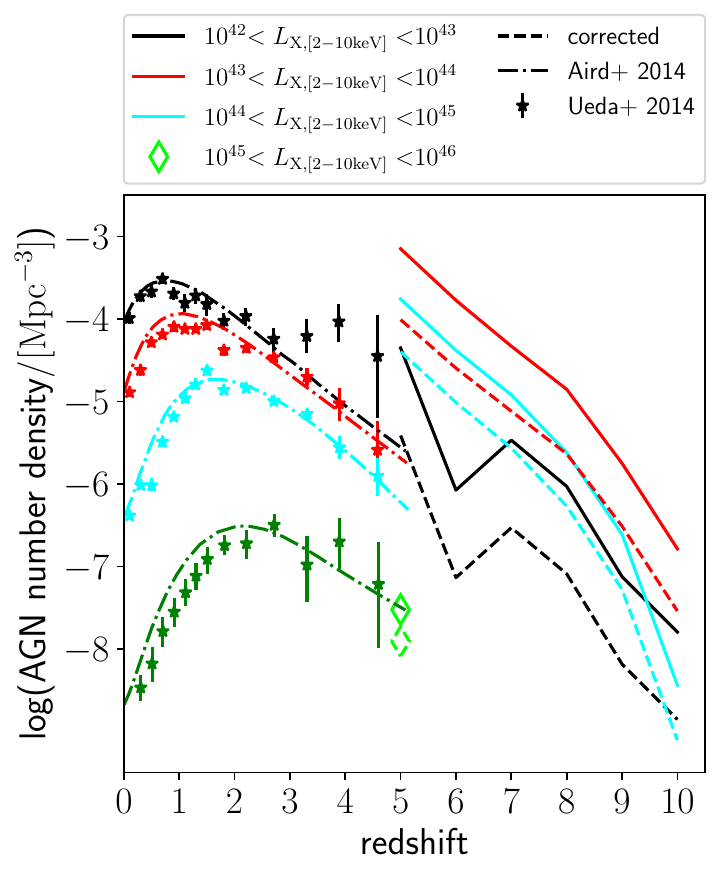}\\
\caption{Evolution of the total AGN intrinsic (solid lines) and obscuration-corrected (dashed lines) number densities for the \textit{fid} model, as a function of hard X-ray luminosity and redshift. In this plot the jet contribution to the X-ray luminosity is not included (i.e. $\mathrm{\alpha_X} = 0$, see text for more details). As our computation extends only to $z \sim 5$, observational results for the models computed for lower $z$ in \protect\citep[][dash-dotted lines, modelled total XLF]{aird2015} and \protect\cite[][stars, modelled Compton-thin AGN XLF only]{ueda2014} are shown for reference. Notice that for $10^{45} \mathrm{erg/s} < L_\mathrm{X} < 10^{46} \mathrm{erg/s}$ our results are shown in light green while the observation-based data in dark green for clarity. The discrepancy between our results and the observation results with assumed absorption model could be related to the uncertainties of obscuration details, as discussed in the text.
}
\label{agnndens}
\end{figure}

In Figure \ref{agnndens} we apply the luminosity-dependent obscuration model from \cite{ueda2014} by computing the unobscured fraction 
\begin{equation}    
f_{\mathrm{unabs}} = \frac{1 - \psi}{1 + \psi},
\end{equation}
where $\psi = 0.43 \left[1 + \min(z, 2) \right]^{0.48} - 0.24 \left(\log L_\mathrm{X}-43.75 \right)$. This function saturates at $0.008$ at the low-luminosity end and at $0.73$ at the high-luminosity end. We then plot both the intrinsic and corrected number densities our \textit{fid} model predicts at $z = 5-10$ for different luminosity bins and in the case $\mathrm{\alpha_X} = 0$. We also show the results from \cite{aird2015} and \cite{ueda2014} - circles and stars - for the redshift range $z = 0-5$. The first work uses surveys with \textit{Chandra}, \textit{ASCA} and \textit{ROSAT} to build AGN samples detected in the soft and hard X-ray bands to separately model the evolution of the the absorbed and unabsorbed AGN X-ray luminosity function to derive the evolution of the total AGN space density (dash-dotted lines). \cite{ueda2014} uses multiple surveys (made with \textit{Swift}/BAT, MAXI, \textit{ASCA}, \textit{XMM-Newton}, \textit{Chandra} and \textit{ROSAT}) to make a population synthesis model, and shows their results for the Compton-thin AGN XLF in Figure \ref{agnndens}. For the lowest and highest luminosity bins ($10^{42} \mathrm{erg/s} < L_\mathrm{X,[2-10\ keV]} < 10^{43} \mathrm{erg/s}$ and $10^{45} \mathrm{erg/s} < L_\mathrm{X,[2-10\ keV]} < 10^{46} \mathrm{erg/s}$) our $z \sim 5$ results are in good agreement with the observations. For the intermediate luminosity bins, our number densities seem up to $1-1.5$ orders of magnitude times higher than those inferred from observations. Parameters tuning can alleviate the discrepancy, but we find it hard to reproduce at the same time the number densities of intermediate and high-luminosity AGN, as we end up overestimating the former or underestimating the latter. Some cautions may be added about this point: first of all, the modelled results from \cite{ueda2014} shown here are valid for the (intrinsic) Compton-thin population, while our intrinsic (i.e.\ uncorrected) X-ray luminosity function takes into account also the contribution from Compton-thick AGN. If their fraction does not exceed $10\;\!\%-20\;\!\%$ as suggested in \cite{ueda2014, aird2015}, taking them into account would not significantly reduce the discrepancy. However the estimates of their number density at such high redshift are still very uncertain. Secondly, the approximations used in the model are expected to affect our results. Our black hole accretion model is aimed at studying the link between the unresolved large-scale gas dynamics of the host galaxy and the time-averaged BH growth rate, but by not considering the microphysics of the BH accretion mechanism we are potentially overlooking short-term variabilities which have an impact on our results for the AGN luminosity functions. Recent X-ray studies showed that variable sources could oscillate in luminosity by up to a factor of 10 or more, showing up in some surveys and not in others \citep{wolf2020, kammoun2023}, and this is likely to occur at high redshift as well. Also for this reason, it is not fully clear how complete and statistically significant the AGN samples of X-ray surveys at high redshift are. One solution could be to try to put together samples from different surveys, as done in both \cite{ueda2014} and \cite{aird2015}. Yet, some part of the most luminous AGN population might still be going undetected, especially at high redshift, both in optical and X-ray surveys \citep[see for instance][]{onken2023, wolf2024}. This might entail changes to the incompleteness corrections that have been applied to previous samples, and, consequently, changes to the observed bright end of the quasar luminosity function. To all these considerations, we should add that in the future it is worth it to test also different AGN obscuration models. 
Some works suggest that the AGN obscured fraction might not depend on the AGN bolometric luminosity, as considered in \cite{ueda2014, merloni2014}, but rather on the Eddington ratio \citep[e.g.][]{ricci2017}. The luminosity-dependent X-ray bolometric correction as well might carry some dependence on the Eddington ratio \citep[see Figure 9 of][]{duras2020}. Finally, we point out that we expect the XLF to fall off towards lower luminosities, as the stellar mass function flattens out at lower stellar masses, while the occupation fraction of X-ray luminous AGN decreases. We predict the XLF to peak at lower luminosities towards higher redshift, and specifically at $L_\mathrm{X} \sim 10^{43.5}$ erg/s at $z=5$ and $L_\mathrm{X} \sim 10^{42.5}$ erg/s at $z=9$, consistently with the evolution of $M_\mathrm{h}^\mathrm{crit}$.

\section{Summary}

In this work, building on the DELPHI cosmological semi-analytic model \citep{dayal2017, piana2021, piana2022}, we explore the cosmic growth of supermassive black holes, and how the evolution of their host galaxies is affected by the black hole accretion and feedback histories. We include gas cooling and heating mechanisms, hot and cold BH accretion, radiative (quasar) and jet (radio) BH feedback modes, and gas re-accretion onto the galaxy, adding them to what remains the key assumption of the model: the critical halo mass threshold $M_\mathrm{h}^\mathrm{crit}$ below which BH growth is hindered by SN feedback \citep{rosasguevara2016, bower2017, lupi2019}. Above this scale, the BH can accrete both from the hot and cold gas phases, respectively through continuous Bondi accretion and through merger-induced accretion episodes. We aim at showing how we can observationally distinguish between different typical BH growth models and at the same time at studying the emergence of jetted AGN at $z>5$. The main results and implications are summarised below. 

\begin{enumerate}
    \item Given our assumptions, the jet-mode feedback is sub-dominant with respect to the radiative feedback, impacting $\approx 1\%-10\;\! \%$ of the total gas mass affected by AGN feedback at $z>4$ (Figure \ref{fig_acch}), but it plays an important role for halos with $M_\mathrm{h} \sim 10^{12} \mathrm{M_\odot}$.
    
    \item By comparing our fiducial super-Eddington ({\it fid}) model to the Eddington-limited ({\it EDDlim}) model, we find the the former predicts a much lower number density of black holes in the range $10^4-10^6 \mathrm{M_\odot}$ at $z>5$. This is due to the fact that BH in that mass range are extremely short-lived, resulting from a very fast growth with $\lambda_\mathrm{E} \sim 10^3$ (Figures \ref{fig_eddrat} and \ref{bhmf}).
    
    \item In both models, the average Eddington ratio tends to decrease as the host galaxy mass increases and the cold gas fraction decreases (Figure \ref{eddr_starm}). This means that active black holes grow faster in higher-z and lower-mass galaxies, and for this reason we observe a bend in the modelled $M_*-M_\mathrm{bh}$ (at $z=5$). At the same time, the normalization of the relationship is slightly higher than the one inferred in the local Universe, consistently with what expected from the redshift evolution of $M_\mathrm{h}^\mathrm{crit}.$
    
    \item The \textit{fid} model predicts BH up to two order of magnitudes more massive than the \textit{EDDlim} model for the same galaxy mass at $z=9$, with BH masses up to $M_\mathrm{bh} \sim 10^{8.5} \mathrm{M_\odot}$ (Figure \ref{bhm_starm}). Moving to lower redshift, at $z=5$, the two models are distinguishable only in terms of the predicted IMBH ($M_\mathrm{bh} \sim 10^{4-6} \mathrm{M_\odot}$) number densities, which are negligible in the super-Eddington scenario. These are clear observational predictions that will allow us to discriminate between a super-Eddington and an Eddington-limited typical BH accretion model once the on-going {\it JWST} surveys and future missions like {\it LISA} will populate the high-$z$ BH-host galaxy correlation planes and BH mass function.
    
    \item Our results for the $M_{bh}-\sigma$ relationship depend on the assumed scaling between the stellar mass and the central velocity dispersion. If we apply the $M_\mathrm{evo}$ fit derived in \cite{cannarozzo2020} for galaxies at $0 \lsim z \lsim 2.5$, we find that our $z \sim 5-9$ results are independent on redshift, and consistent with local observations (Figure \ref{sigma_bhm}). Given that the fitted sample was formed by ETGs, we can suppose that it should be valid for our most massive galaxies if we assume these to be the direct progenitors of the low-$z$ ETGs.

    \item Given our AGN jet model, we show that by $z \sim 5$ SMBHs with $M_\mathrm{bh} \sim 10^6 \mathrm{M_\odot}$ have spent close to $95\%$ of their lifetime in jetted mode, while the most massive black holes only $20\%$, independently on the time resolution of our merger tree (Figure \ref{sdt}).
    
    \item We have found that the current model that we have adopted either overpredicts the AGN number densities for AGNs with $10^{43} \mathrm{erg/s} < L_\mathrm{X,[2-10\ keV]} < 10^{45} \mathrm{erg/s}$ at $z=5$ (Figure \ref{agnndens}) or underpredicts the number densities of high-luminosity AGN. Nevertheless, if we compare our XLF with the results at $z \sim 7$ from other theoretical models and numerical simulations, we find a relative good agreement (Figure \ref{xlf}). In our $fid$ model, in which lower-mass black holes can go through strong hyper-Eddington accretion phases, we observe a generally flatter XLF than in models implementing Eddington-limited accretion rates. .
\end{enumerate}

Finally, it is worth touching upon a few other caveats. We are not taking into account any metallicity evolution, which directly impacts the gas cooling rates in the galaxy, and if it is true that low-mass galaxies at high redshift have been consistently observed with low metallicity, this has not been the case for high-redshift AGN \citep{decarli2018, venemans2019}. In particular, higher metallicities means softer spectra, which in turn means that higher stellar masses are required to reproduce the same stellar UV luminosity function. When considering black hole growth, though, AGN luminosity depends only on the amount of gas accreted, not on its metallicity. In this sense, black hole activity is self-regulated: higher accretion rates from higher cooling rates will eventually lead to more feedback and more ejected gas mass, and hence less fuel and lower accretion rates in the subsequent time step. Therefore, we do not expect the metallicity to significantly affect our resulting black hole masses. We will further study and discuss this in a future work, where we plan to implement a consistent metallicity treatment and to extend the model to $z<4$. Secondly, we are neglecting stellar mass losses due to winds, which would return part of the stellar mass to the gas phase and inject more energy into the ISM. However, the energy injection is dominated by high-mass stars, but is still subdominant with respect to SN feedback. At the same time, the gas mass returned to the ISM through winds is mostly contributed by lower-mass stars, which do not evolve much during the first Gyr of life of the Universe. In addition, our transitional halo mass $M_\mathrm{h}^\mathrm{crit}$ is unlikely to be a universal value, and should depend on the environment in which galaxies sit, and on the reionization state of the Universe. When taken into account, we would expect some BH to start growing in lower-mass galaxies and some in higher-mass galaxies, hence the IMBH gap visible in the BH mass function in the \textit{fid} case would partially fill up. In addition, \cite{ocvirk2008} showed that at high redshift the fraction of gas $f_\mathrm{cold}$ accreted in cold phase by the galaxy is not a step function of the halo mass, as we assume here, rather a smooth decreasing function. However, when we take this effect into account we notice no significant change in our results. Indeed, the self-regulating action of BH growth and feedback in high-mass halos, as shown in Figure \ref{fig_eddrat}, smooths out the difference between the two implementations. We plan to address these issues in future works, as we expand the model down to $z=0$. 

\section*{Acknowledgments} 
We thank Christian Wolf for a useful discussion, which led to improvement of the manuscript, and the anonymous referee for their insightful comments. OP thanks professor Antonaldo Diaferio and the Department of Physics of the University of Torino for hosting him during the first months of this project. OP and HYP are supported by the Yushan Young Scholar Program of the Ministry of Education (MoE) of Taiwan (ROC). HYP is also supported by the National Science and Technology Council (NSTC) of Taiwan (ROC) under the grant 112-2112-M-003-010-MY3. KW acknowledges the support from the UCL Cosmoparticle Initiative. This work has made use of the NASA Astrophysics Data System. 

\section*{Data Availability}
The data underlying this paper, which have been produced by our model, are available on reasonable request to the corresponding author.

\bibliographystyle{mn2e}
\bibliography{lib}

\label{lastpage} 
\end{document}